\documentclass[12pt]{article}
\usepackage{amssymb}
\usepackage{graphicx}

\begin{document}

\begin{titlepage}
\begin{center}

\vspace*{10mm}

{\LARGE\bf
A closer look at string resonances in dijet events at the LHC}
 
\vspace*{20mm}

{\large
Noriaki Kitazawa
}
\vspace{6mm}

{\it
Department of Physics, Tokyo Metropolitan University,\\
Hachioji, Tokyo 192-0397, Japan\\
e-mail: kitazawa@phys.metro-u.ac.jp
}

\vspace*{15mm}

\begin{abstract}
The first string excited state can be observed as a resonance in dijet
invariant mass distributions at the LHC, if the scenario of low-scale string
with large extra dimensions is realized. A distinguished property of the
dijet resonance by string excited states from that the other ``new physics''
is that many almost degenerate states with various spin compose a single
resonance structure. It is examined that how we can obtain evidences of
low-scale string models through the analysis of angular distributions of
dijet events at the LHC. Some string resonance states of color singlet can
obtain large mass shifts through the open string one-loop effect, or through
the mixing with closed string states, and the shape of resonance structure
can be distorted. Although the distortion is not very large (10\% for the
mass squared), it might be able to observe the effect at the LHC, if gluon
jets and quark jets could be distinguished in a certain level of efficiency.
\end{abstract}

\end{center}
\end{titlepage}

\section{Introduction}
\label{introduction}

The low-scale string scenario \cite{Antoniadis:1990ew}
(and see Refs.\cite{Antoniadis:2004wm,Antoniadis:2007uz} for review)
is an interesting possibility which may be confirmed or excluded by
the LHC. Low-energy supersymmetry is not necessary and electroweak
symmetry breaking can be triggered by one-loop string effects
\cite{Antoniadis:2000tq,Kitazawa:2006if}. The string scale
($M_s \equiv 1/\sqrt{\alpha'}$) should be at least less than about $10$TeV
for the right scale of electroweak symmetry breaking. 

The discovery reach through the observation of resonances in dijet
events by first string excited states at the LHC has been investigated in 
\cite{Cullen:2000ef,Anchordoqui:2008di,Anchordoqui:2008hi,Anchordoqui:2009mm}.
The important point is that the processes between gluons and quarks,
$gg \rightarrow gg$, $qg \rightarrow qg$, ${\bar q}g \rightarrow {\bar qg}$,
$gg \rightarrow q {\bar q}$ and $q {\bar q} \rightarrow gg$,
can be described in String Theory in a model independent way.
The simple fact that the gauge symmetry $SU(3)_c$ for the strong interaction
is on one D-brane stack can almost completely determine the amplitudes of these
processes, especially for the process $gg \rightarrow gg$.

In case of $M_s=2$ TeV, for example, a resonance structure will appear at the
invariant mass $M=2$ TeV in dijet invariant mass distribution at the LHC.
Among many other possible ``new physics'', like quark compositeness,
a distinguished property of string models is that the resonance structure
consists of many independent states with almost degenerate masses. There are
following six states which contribute to one resonance structure:
color octets with $J=0$ and $2$, color singlets with $J=0$ and $2$, and color
triplets with $J=1/2$ and $3/2$. The origin of color singlet states is the fact
that on a $n$ multiple D-brane a simple gauge symmetry SU$(n)$ is not realized
but a semi-simple U$(n)=$ SU$(n)\times$U$(1)$ is realized. This additional
U$(1)$ results color singlet excited states in the above processes, except for
$qg \rightarrow qg$ and ${\bar q}g \rightarrow {\bar qg}$, where color
triplet states contribute.

It is very important to distinguish observed resonance by string excited
states from that by other possible ``new physics''. In this article we propose
two way of analysis. One is revealing the multi-states contribution by dijet
angular analysis, and the other is confirming the existence of color singlet
states through the distinction between gluon jets and quark jets (though it 
would be very difficult experimentally). 

Since the process $qg \rightarrow qg$ dominates the others in almost all the
cases, angular distributions can be used to reveal the contribution of
highly degenerate two states, color triplets with $J=1/2$ and $3/2$,
in that process. Our angular analysis is not the same that has been done in
ref.\cite{Anchordoqui:2009mm}, but that has been done experimentally at the
Tevatron (see \cite{Nunnemann:2009py} and references therein). This angular
analysis is given in section \ref{angular} after a review on string amplitudes
and dijet evens in the next section. In section \ref{angular} we show the
possibility of revealing two state contribution through the fitting of
angular distribution using the expected behavior from $J=1/2$ and $3/2$ states.

It is possible that color singlet states obtain large mass shifts through
the mixing with closed string states. The similar effect happens in the
generation of the masses of anomalous U$(1)$ gauge bosons in D-brane
string models. In section \ref{mass-shift} we give a brief estimate of
the magnitude of the mass shift through the calculation of open string
one-loop diagrams, and obtain the result that about $10\%$ shifts in mass
squared exist. In section \ref{gluon-dijet} we discuss the distortion of
the shape of the resonance structure in {\it gluon} dijet invariant mass
distributions. We also discuss realistic situation with the dominant process
$qg \rightarrow qg$ and show the possibility to detect the distortion.

In the last section we give a short summary of our results and point out
necessary future works.

\section{String amplitudes and cross sections for dijet evens}
\label{amplitudes}

String amplitudes (amplitudes calculated in string world-sheet theory)
of two body scattering processes between gluons and quarks are calculated
in ref.\cite{Lust:2008qc} and summarized in ref.\cite{Anchordoqui:2009mm}.
The squared amplitude with initial polarization averaged and
finial polarization summed are given as follows.
\begin{eqnarray}
\vert {\cal M}(gg \rightarrow gg) \vert^2
 &=&
 {{19} \over {12}} {{g^4} \over {M_s^4}}
 \Bigg\{
  {{25} \over {57}}
  \left[
   {{M_s^8} \over {(\hat{s}-M_s^2)^2 + (M_s \Gamma_{g^*}^{J=0})^2}}
  +{{\hat{t}^4 + \hat{u}^4} \over
    {(\hat{s}-M_s^2)^2 + (M_s \Gamma_{g^*}^{J=2})^2}}
  \right]
\label{amp2-gg}
\nonumber\\
 && \quad +
  {{32} \over {57}}
  \left[
   {{M_s^8} \over {(\hat{s}-M_s^2)^2 + (M_s \Gamma_{C^*}^{J=0})^2}}
  +{{\hat{t}^4 + \hat{u}^4} \over
    {(\hat{s}-M_s^2)^2 + (M_s \Gamma_{C^*}^{J=2})^2}}
  \right]
 \Bigg\},
\\
\vert {\cal M}(gg \rightarrow q {\bar q}) \vert^2
 &=&
 {7 \over {24}}{{g^4} \over {M_s^4}} N_f
 \left[
  {5 \over 7}
  {{\hat{u}\hat{t} (\hat{u}^2 + \hat{t}^2)}
   \over
   {(\hat{s}-M_s^2)^2 + (M_s \Gamma_{g^*}^{J=2})^2}}
 +{2 \over 7}
  {{\hat{u}\hat{t} (\hat{u}^2 + \hat{t}^2)}
   \over
   {(\hat{s}-M_s^2)^2 + (M_s \Gamma_{C^*}^{J=2})^2}}
 \right],
\\
\vert {\cal M}(q {\bar q} \rightarrow gg) \vert^2
 &=&
 {{56} \over {27}}{{g^4} \over {M_s^4}}
 \left[
  {5 \over 7}
  {{\hat{u}\hat{t} (\hat{u}^2 + \hat{t}^2)}
   \over
   {(\hat{s}-M_s^2)^2 + (M_s \Gamma_{g^*}^{J=2})^2}}
 +{2 \over 7}
  {{\hat{u}\hat{t} (\hat{u}^2 + \hat{t}^2)}
   \over
   {(\hat{s}-M_s^2)^2 + (M_s \Gamma_{C^*}^{J=2})^2}}
 \right],
\\
\vert {\cal M}(qg \rightarrow qg) \vert^2
 &=& \vert {\cal M}({\bar q}g \rightarrow {\bar q}g) \vert^2
\nonumber\\
 &=&
 {4 \over 9}{{g^4} \over {M_s^2}}
 \left[
  {{M_s^4(-\hat{u})}
   \over
   {(\hat{s}-M_s^2)^2 + (M_s \Gamma_{q^*}^{J=1/2})^2}}
 +{{(-\hat{u})^3}
   \over
   {(\hat{s}-M_s^2)^2 + (M_s \Gamma_{q^*}^{J=3/2})^2}}
 \right],
\\
\vert {\cal M}(gq \rightarrow gq) \vert^2
 &=& \vert {\cal M}(g{\bar q} \rightarrow g{\bar q}) \vert^2
\nonumber\\
 &=&
 {4 \over 9}{{g^4} \over {M_s^2}}
 \left[
  {{M_s^4(-\hat{t})}
   \over
   {(\hat{s}-M_s^2)^2 + (M_s \Gamma_{q^*}^{J=1/2})^2}}
 +{{(-\hat{t})^3}
   \over
   {(\hat{s}-M_s^2)^2 + (M_s \Gamma_{q^*}^{J=3/2})^2}}
 \right],
\end{eqnarray}
 where $\hat{s}$, $\hat{t}$ and $\hat{u}$ are Mandelstam variables
 for partons, $g = \sqrt{4 \pi \alpha_s}$ is the gauge coupling of the
 strong interaction, and $N_f=6$ is the number of quark flavors.

The decay widths of each string resonances (color octets $g^*$, color
singlets $C^*$, and color triplets $q^*$) are calculated from string
amplitudes in \cite{Anchordoqui:2008hi}.
\begin{eqnarray}
 \Gamma_{g^*}^{J=0} &=& {{N_c} \over 4} \alpha_s M_s
                    \simeq 75 \left( {{M_s} \over {1000 {\rm GeV}}} \right)
                    {\rm GeV},
\\
 \Gamma_{C^*}^{J=0} &=& {{N_c} \over 2} \alpha_s M_s
                    \simeq 150 \left( {{M_s} \over {1000 {\rm GeV}}} \right)
                    {\rm GeV},
\\
 \Gamma_{g^*}^{J=2} &=& \left(
                         {{N_c} \over {10}}
                         + {{N_f} \over {40}}
                        \right)  \alpha_s M_s
                    \simeq 45 \left( {{M_s} \over {1000 {\rm GeV}}} \right)
                    {\rm GeV},
\\
 \Gamma_{C^*}^{J=2} &=& \left(
                         {{N_c} \over 5}
                         + {{N_f} \over {40}}
                        \right)  \alpha_s M_s
                    \simeq 75 \left( {{M_s} \over {1000 {\rm GeV}}} \right)
                    {\rm GeV},
\\
 \Gamma_{q^*}^{J=1/2} &=& \Gamma_{q^*}^{J=3/2} =
                      {{N_c} \over 8} \alpha_s M_s
                    \simeq 38 \left( {{M_s} \over {1000 {\rm GeV}}} \right)
                    {\rm GeV},
\end{eqnarray}
 where $N_c=3$ is the number of color.
Four first string excited states $g^*$'s and $C^*$'s can decay into lowest
lying state corresponding to U$(1)_c$ gauge boson in the gauge symmetry
U$(3)_c$=SU$(3)_c\times$U$(1)_c$ on a ``color D-brane''. The stats is not a
mass eigenstate, but a certain model dependent combination of anomalous U$(1)$
gauge bosons and hypercharge gauge boson, and their masses are model dependent
(heavier than about $0.1 \times M_s^2$ except for hypercharge gauge boson).
In the calculation of ref.\cite{Anchordoqui:2008hi} it is assumed that U$(1)_c$
gauge boson is a massless eigenstate for simplicity, and the real widths for
the states $g^*$'s and $C^*$'s could be smaller.

The differential cross sections at the parton level are given by
\begin{equation}
 {{d\sigma} \over {d\hat{t}}}
  = {{\vert {\cal M} (\hat{s},\hat{t},\hat{u}) \vert^2}
     \over {16 \pi \hat{s}^2}}.
\end{equation}
Note that six first string excited states have the same mass $M_s$ and can
give Breit-Wigner type resonances at the same place $\hat{s} = M_s^2$.

Since the LHC is proton-proton collider, center of mass energies of colliding
partons do not have fixed values, but follow certain distributions described
by parton distribution functions. An observable quantities is the dijet
invariant mass distribution defined for the process,$ij \rightarrow kl$,
for example, as
\begin{equation}
 {{d \sigma_{ij \rightarrow kl}} \over {dM}}
 = M
   \int dY \ x_1 f_i(x_1,M^2) \ x_2 f_j(x_2,M^2)
   \int dy \ {1 \over {\cosh^2(y)}}
             \cdot {{d \sigma_{ij \rightarrow kl}} \over {d \hat{t}}},
\label{inv-mass-dist}
\end{equation}
where $f_i(x,M^2)$ is the parton distribution function of parton $i$
(we use CTEQ6D parton distribution functions \cite{Pumplin:2002vw}),
and $M$ is the invariant mass of two jets originated from partons
$k$ and $l$, or ideally $M^2=(p_k+p_l)^2$. Integrations over Bjorken's $x$,
$x_1$ and $x_2$, are reparameterized as integrations over rapidities,
$Y$ and $y$, where $Y \equiv (y_1 + y_2)/2$ describes the amount of boost
and $y \equiv (y_1 - y_2)/2$ describes the angular distribution up to the
boost with $y_1$ and $y_2$ are rapidities of the jets originated from
$k$ and $l$, respectively.
The quantities $x_1$, $x_2$, $\hat{s}$, $\hat{t}$ and $\hat{u}$ are described
as follows.
\begin{equation}
 x_1 = \sqrt{{{M^2} \over s}} e^Y,
 \qquad
 x_2 = \sqrt{{{M^2} \over s}} e^{-Y}
\end{equation}
 with $\sqrt{s}$ is the center of mass energy of $pp$ collision,
 and
\begin{equation}
 \hat{s} = M^2,
\quad
 \hat{t} = - {{M^2} \over 2} {{e^{-y}} \over {\cosh(y)}},
\quad
 \hat{u} = - {{M^2} \over 2} {{e^y} \over {\cosh(y)}}.
\end{equation}
Experimentally, we need to set the maximal absolute value of rapidities,
$\vert y_1 \vert, \vert y_2 \vert < 1$, for example. In this case,
$y_{\rm max} = 1$ and the integration region becomes
\begin{equation}
 \int_{-1}^1 dy_1 \int_{-1}^1 dy_2
 = \int_{-Y_{\rm max}}^0 dY \int_{-(y_{\rm max} + Y)}^{y_{\rm max} + Y} dy
 + \int_0^{Y_{\rm max}} dY \int_{-(y_{\rm max} - Y)}^{y_{\rm max} - Y} dy
\end{equation}
 with $Y_{\rm max} = {\rm min}(\ln(\sqrt{s/M^2}), y_{\rm max})$.

\begin{figure}[t]
\centering
\includegraphics[width=45mm]{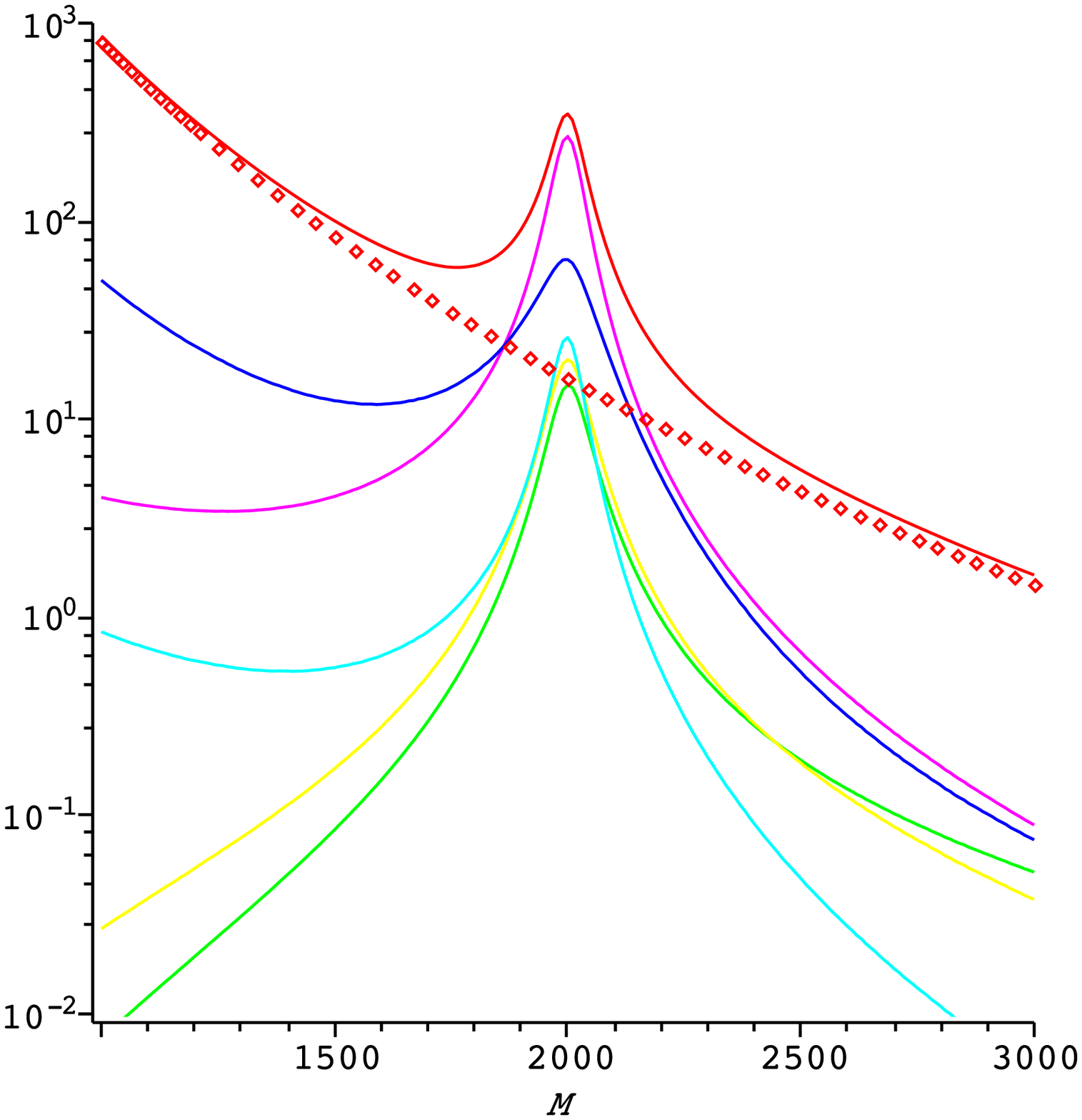}
\includegraphics[width=45mm]{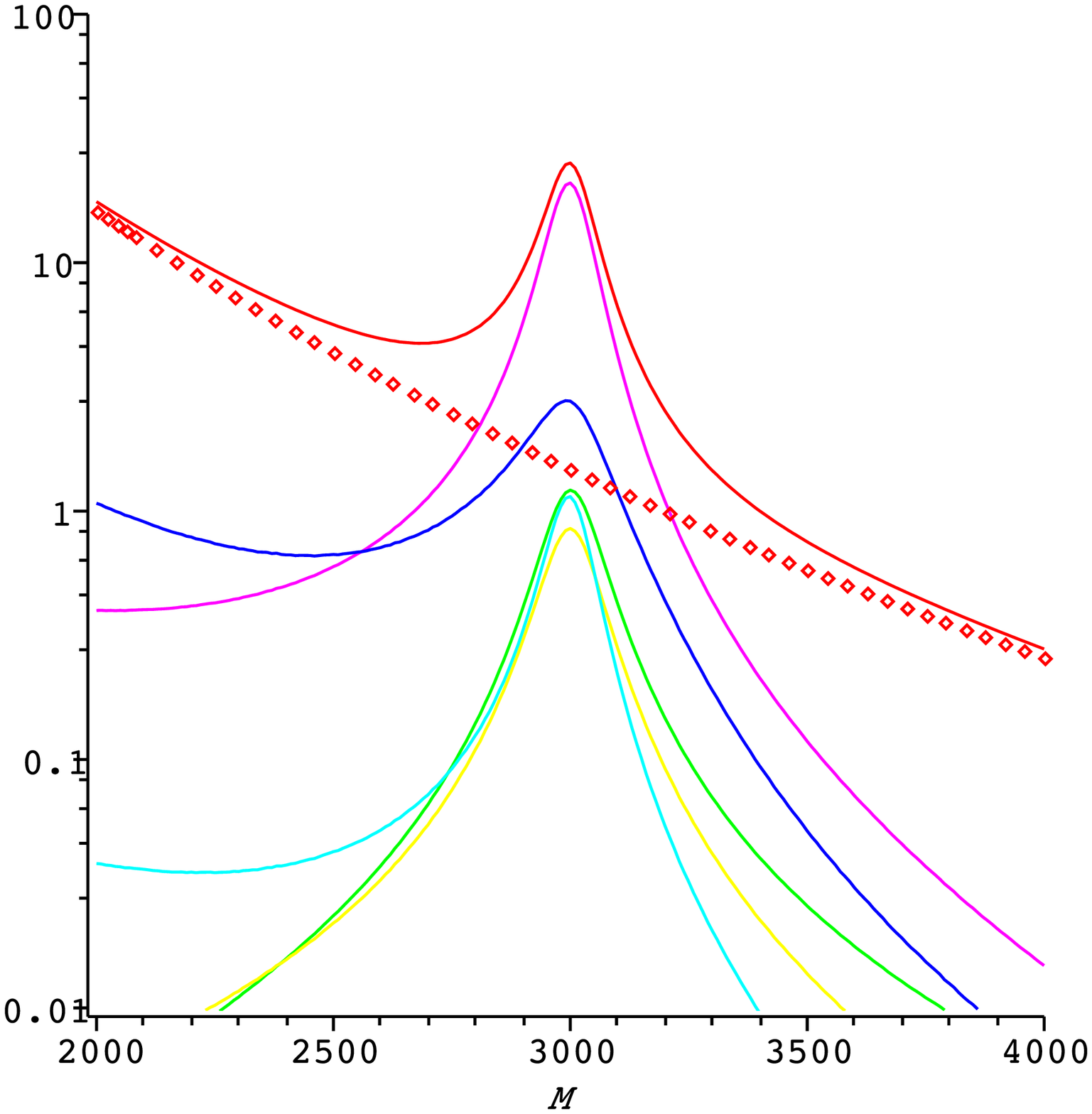}
\includegraphics[width=45mm]{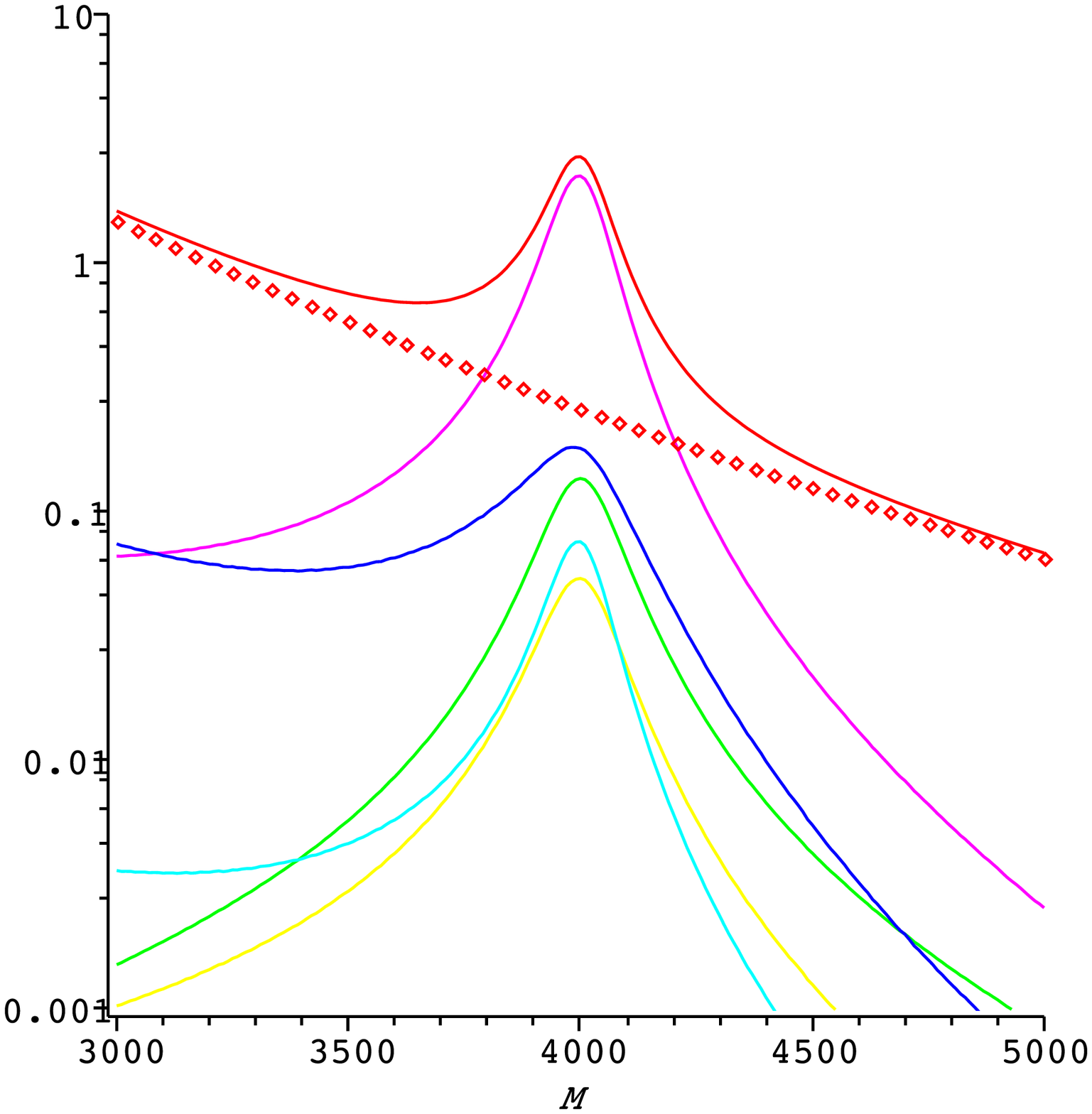}
\caption{
Invariant mass distributions $d\sigma/dM$ [fb/GeV] for the case of $M_s=2,3,4$
TeV from left to right.
The correspondence of the colored line to the processes as,
blue: $gg \rightarrow gg$, magenta: $qg \rightarrow qg$,
cyan: ${\bar q}g \rightarrow {\bar q}g$, yellow: $gg \rightarrow q {\bar q}$,
green: $q {\bar q} \rightarrow gg$,
red: QCD background plus string resonances,
and red points: QCD background.
}
\label{fig:resonances-all}
\end{figure}

The invariant mass distributions in case of $M_s=2,3$ and $4$ TeV are given in
Fig.\ref{fig:resonances-all} with $\vert y_1 \vert, \vert y_2 \vert < 1$
and $\sqrt{s}=14$ TeV.
We see that the process $qg \rightarrow qg$ always dominates at the place
of the resonances. This is due to the effects of parton distribution functions
and of small width of color triplet excited states.

\section{Angular analysis of dijet events}
\label{angular}

It is natural to consider angular distribution of dijets to reveal that
resonance structure in dijet invariant mass distribution consists many states
with different spins. We consider $\chi \equiv \exp(2y)$ distributions
with $M^2$ integrated over some certain ranges (see \cite{Nunnemann:2009py}
and references therein). The $\chi$-distribution is defined from
eq.(\ref{inv-mass-dist}) as
\begin{equation}
 {{d \sigma_{ij \rightarrow kl}} \over {d \chi}}
 = \int_{M^2_{\rm low}}^{M^2_{\rm high}} dM^2
   \int_{-Y_{max}}^{Y_{max}} dY \ x_1 f_i(x_1,M^2) \ x_2 f_j(x_2,M^2) \
    {1 \over {(1+\chi)^2}}
    \cdot {{d \sigma_{ij \rightarrow kl}} \over {d \hat{t}}}.
\label{chi-dist}
\end{equation}
In the following we take $Y_{\rm max}=1$ and $1 < \chi < 20$ which corresponds
to $0 < y < 1.5$ (and $\sqrt{s}=14$ TeV). We consider two kinds of integration
regions on $M^2$:
\begin{eqnarray}
 M^2_{\rm low} = (M_s-250 \, \mbox{GeV})^2,
 \quad M^2_{\rm high} = (M_s+250 \, \mbox{GeV})^2
 \qquad \mbox{: ``on peak''}
\\
 M^2_{\rm low} = (M_s-750 \, \mbox{GeV})^2,
 \quad M^2_{\rm high} = (M_s-250 \, \mbox{GeV})^2
 \qquad \mbox{: ``off peak''}
\end{eqnarray}
As shown in Fig.\ref{fig:resonances-all} $qg \rightarrow qg$ process
dominates for ``on peak'' case in any value of $M_s = 2,3$ and $4$. It is
also true for ``off peak'' case except for the case of $M_s = 2$, where
$gg \rightarrow gg$ process dominates.

\begin{figure}[t]
\centering
\includegraphics[width=45mm]{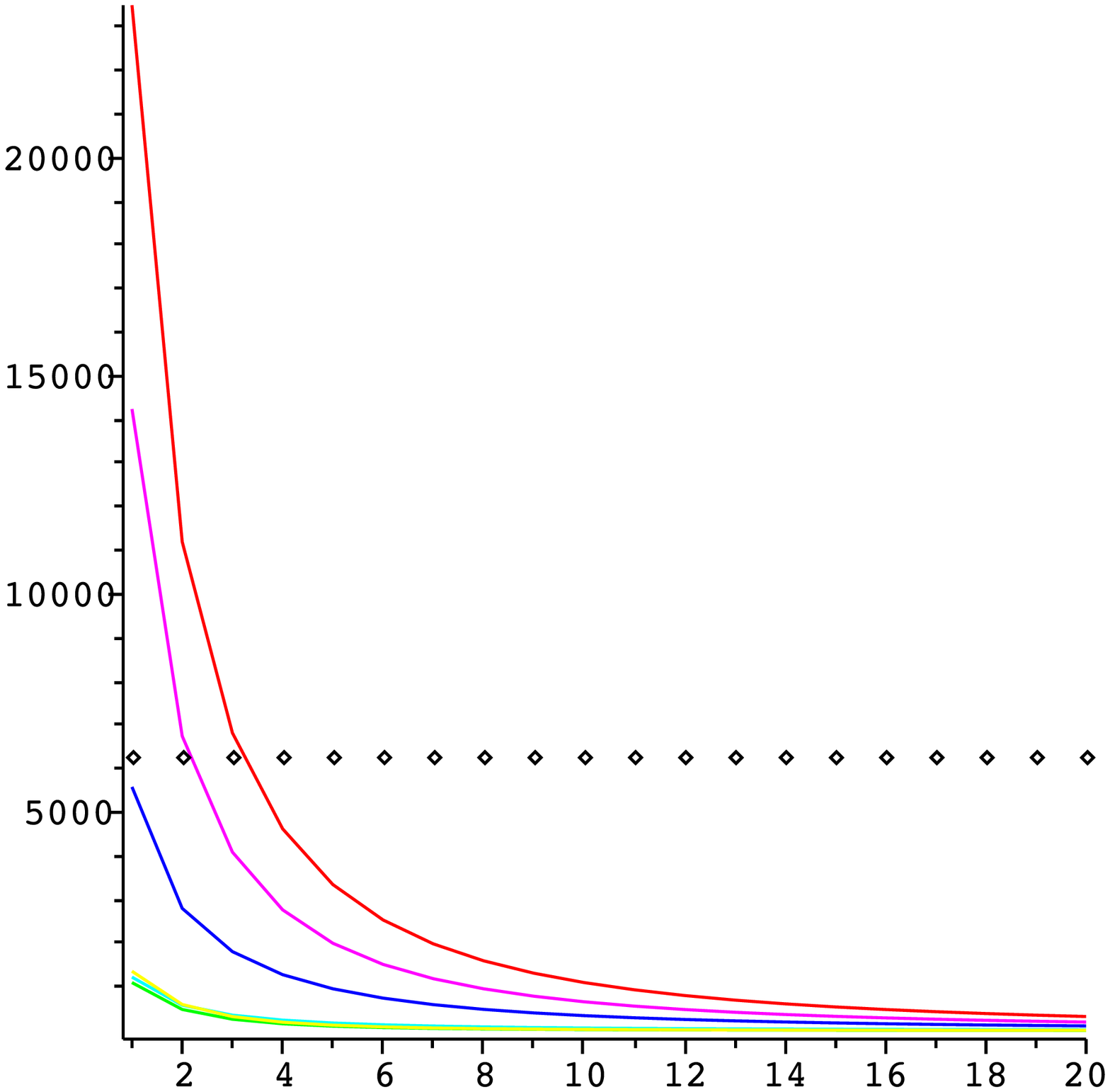}
\includegraphics[width=45mm]{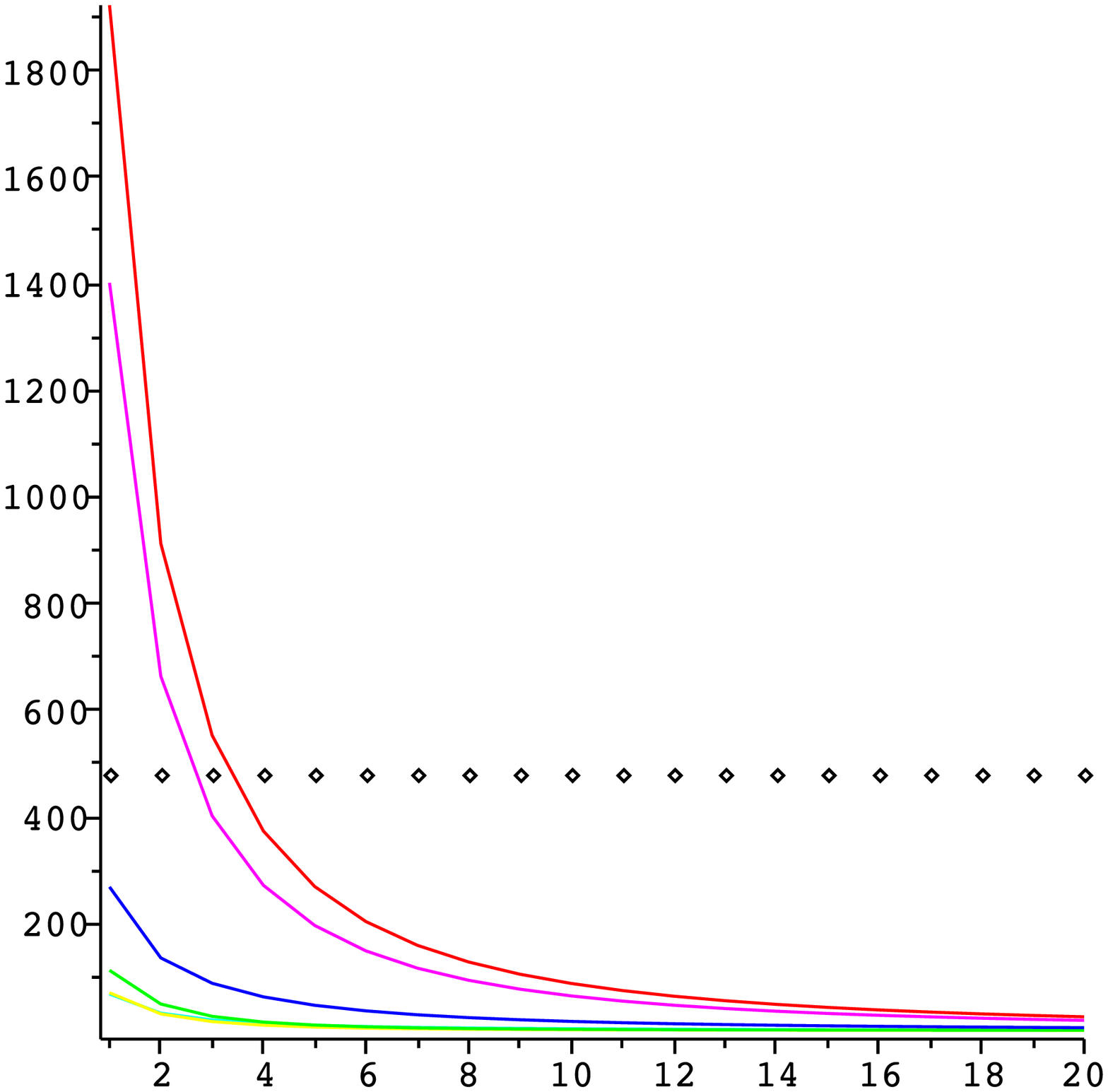}
\includegraphics[width=45mm]{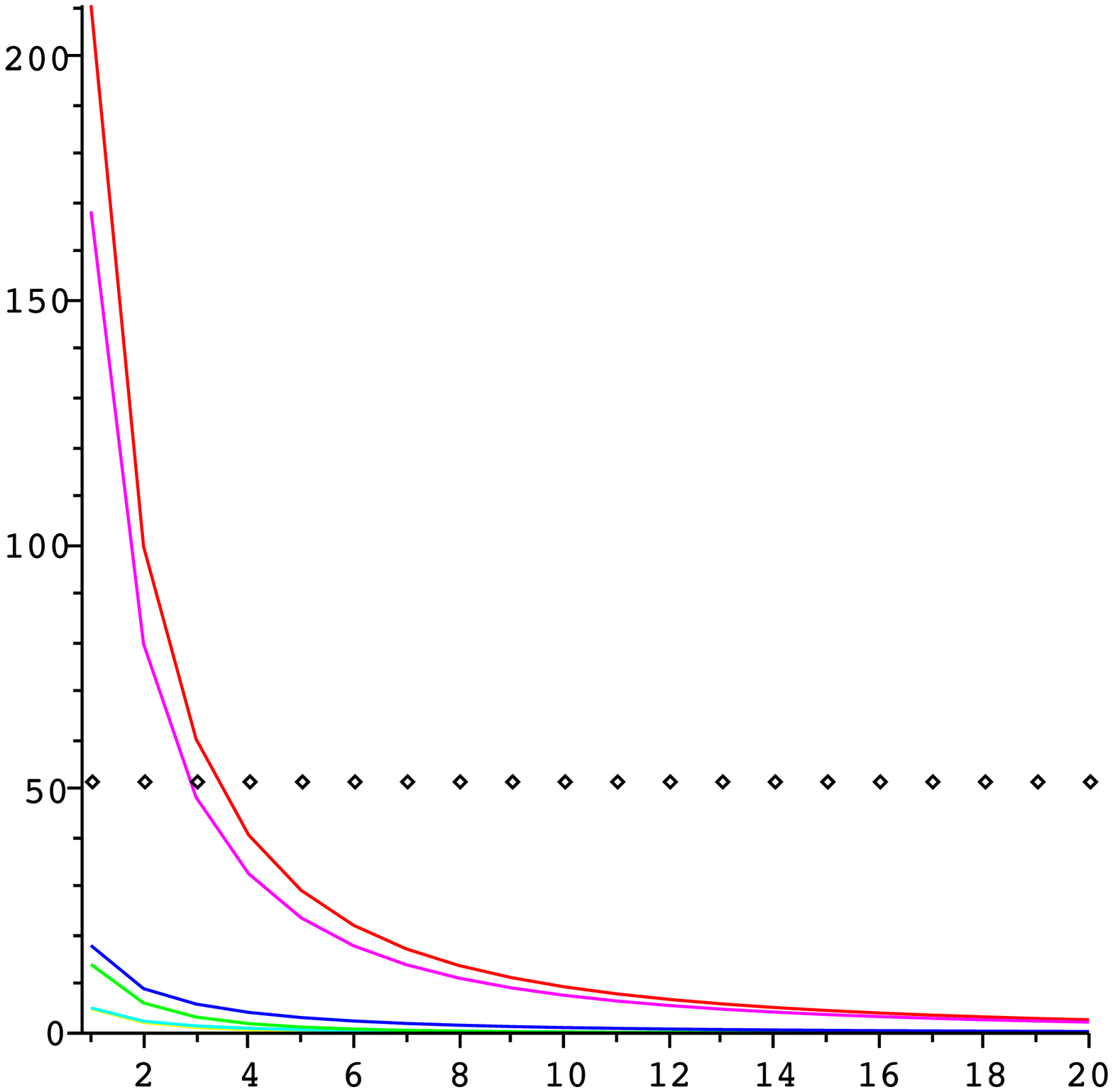}
\caption{
``on peak'' $\chi$-distributions $d\sigma/d\chi$ [fb] for the case of
$M_s=2,3$ and $4$ TeV from left to right. The color mapping to six processes
are the same in Fig.\ref{fig:resonances-all} except for that red line indicates
total distribution. Black dots indicate rough estimates of QCD backgrounds.
}
\label{chi-separate-on}
\end{figure}

\begin{figure}[t]
\centering
\includegraphics[width=45mm]{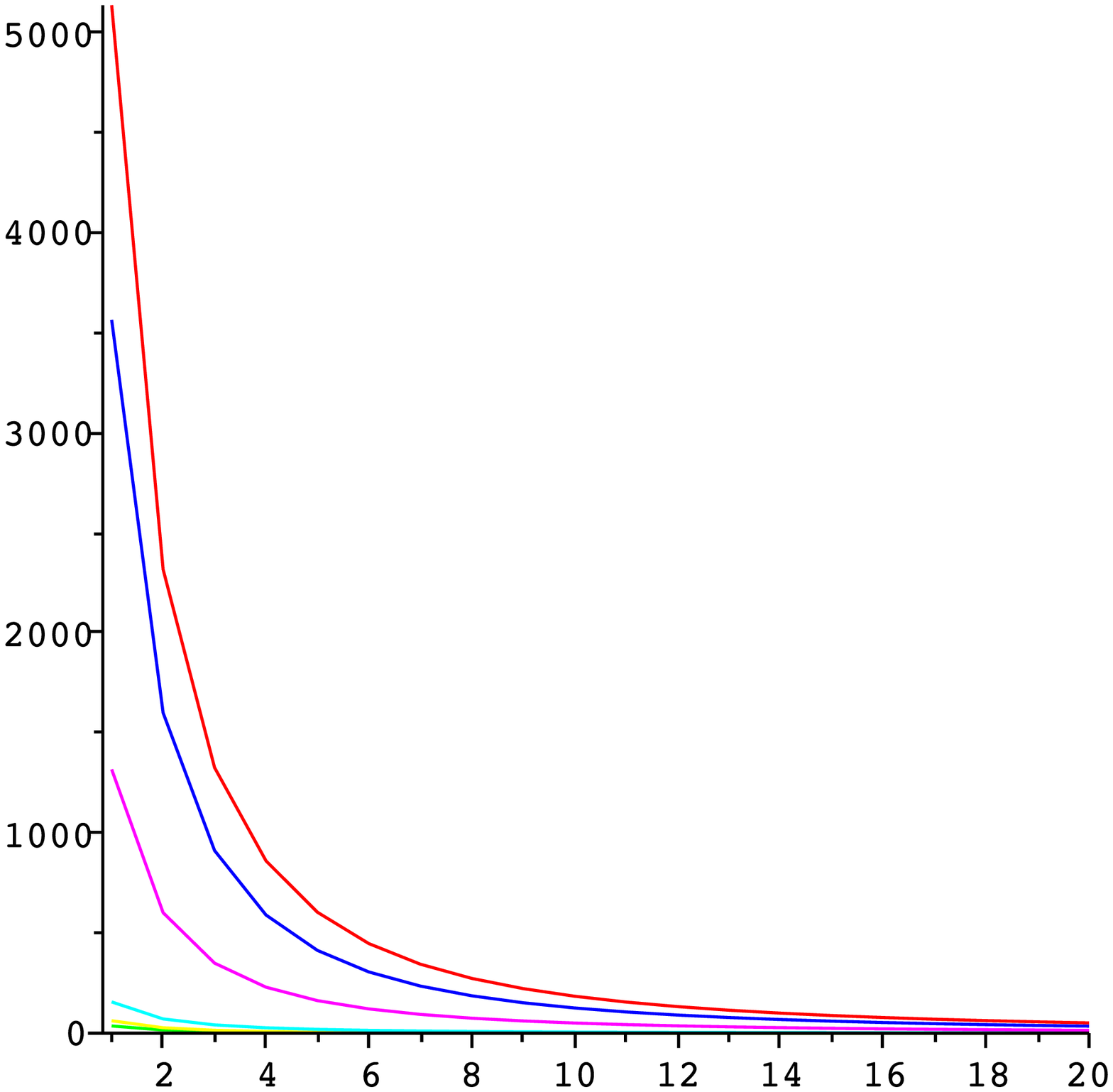}
\includegraphics[width=45mm]{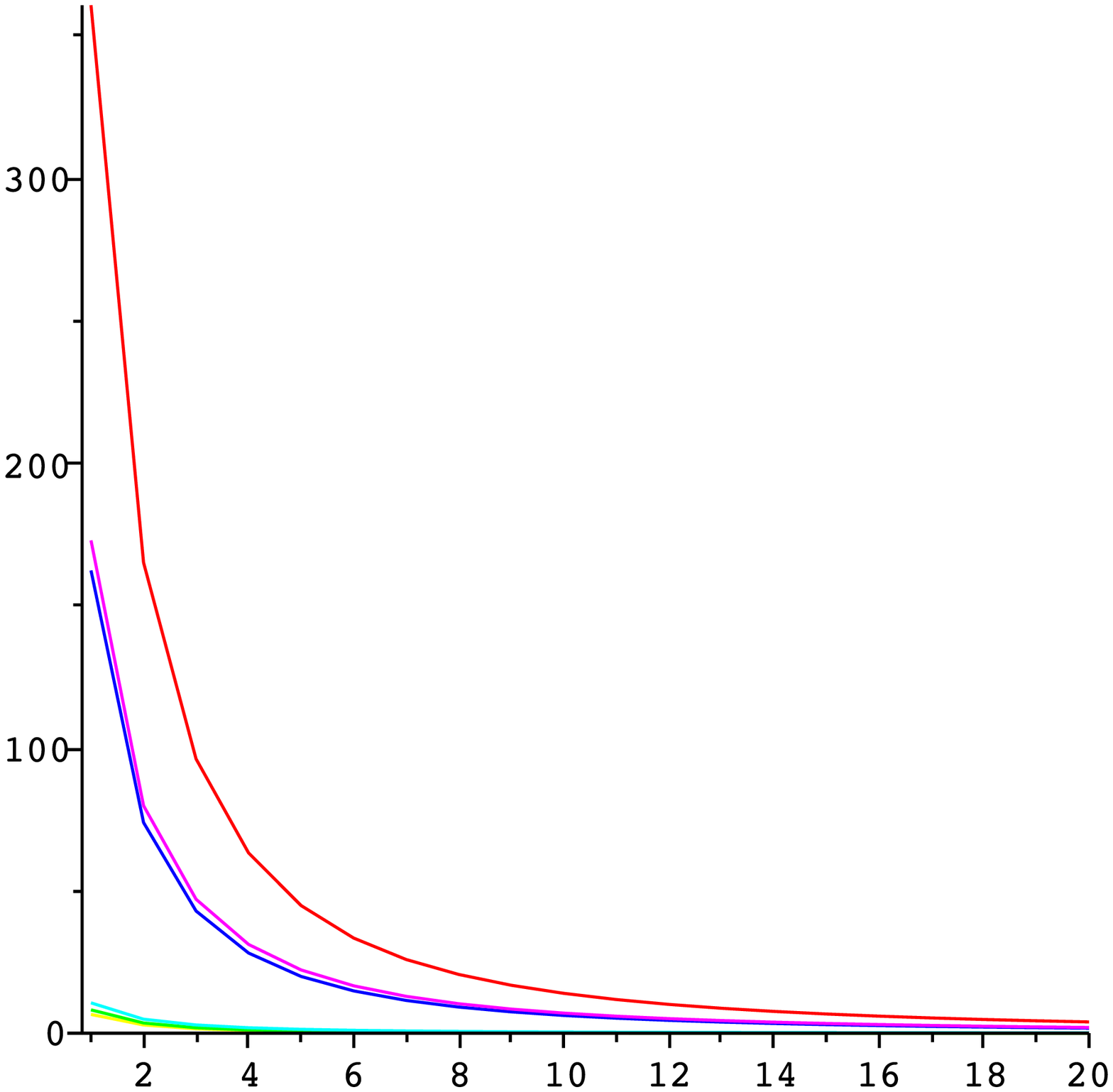}
\includegraphics[width=45mm]{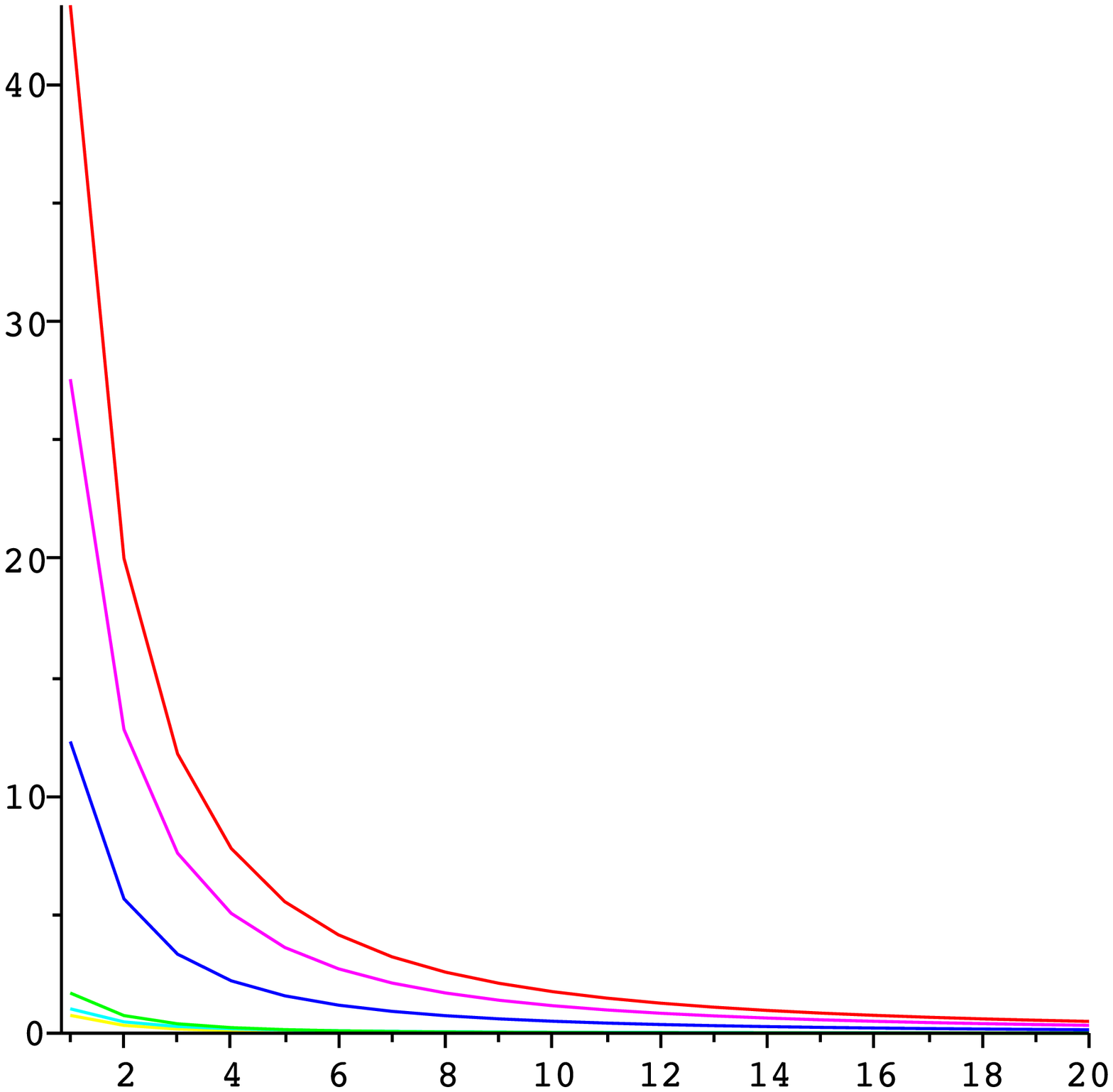}
\caption{
``off peak'' $\chi$-distributions $d\sigma/d\chi$ [fb] for the case of
$M_s=2,3$ and $4$ TeV from left to right. The color mapping to six processes
are the same in Fig.\ref{fig:resonances-all} except for that red line indicates
total distribution.
}
\label{chi-separate-off}
\end{figure}

Fig.\ref{chi-separate-on} show individual contributions of six processes to
``on peak'' $\chi$-distribution and Fig.\ref{chi-separate-off} show the
same for ``off peak'' case.
It is expected that the distribution of QCD background is almost flat,
and large values at small $\chi$ indicate ``new physics''
(for detail see Ref.\cite{Boelaert:2009jm}, for example).
Fig.\ref{chi-separate-on} includes very roughly estimated QCD backgrounds.
For ``off peak'' case QCD background is very large:
about $29000$ [fb], $1600$ [fb] and $150$ [fb] for $M_s=2,3$ and $4$ TeV,
respectively.
In case of ``on peak'' the process of $qg \rightarrow qg$ always dominates
and other processes are negligible at large values of $M_s \gtrsim 4$ TeV.
In case of ``off peak'' the process of $gg \rightarrow gg$ dominates
at small value of $M_s \sim 2$ TeV due to the effect of gluon parton
distribution function. At higher value of $M_s > 3$, the process of
$qg \rightarrow qg$ dominates the others also in ``off peak'' case.

Six processes predict the following form of $\chi$-distributions.
\begin{eqnarray}
 && {1 \over {(1+\chi)^2}}
  \left(A_1 + B_1 \ {{1+\chi^4} \over {(1+\chi)^4}} \right)
  \quad \mbox{for $gg \rightarrow gg$},
\label{J=0-2}\\
 && {1 \over {(1+\chi)^2}}
  \left(A_2 + B_2 \ {{1+\chi^3} \over {(1+\chi)^3}} \right)
  \quad \mbox{for $qg \rightarrow qg$ and ${\bar q}g \rightarrow {\bar q}g$},
\label{J=1/2-3/2}\\
 && {1 \over {(1+\chi)^2}}
  \ B_3 \ {{\chi(1+\chi^3)} \over {(1+\chi)^4}}
  \quad \mbox{for $gg \rightarrow q{\bar q}$ and $q{\bar q} \rightarrow gg$},
\label{J=1in2}
\end{eqnarray}
 where $A_1, A_2, B_1, B_2$ and $B_3$ are constants.
The common factor $1/(1+\chi)^2$ is a kinematical one in eq.(\ref{chi-dist}).
The first distribution, eq.(\ref{J=0-2}), indicates a combination of spin $0$
and spin $2$ intermediate states with two massless spin $1$ gluons as initial
states.
The second distribution, eq.(\ref{J=1/2-3/2}), indicates a combination of
spin $1/2$ and $3/2$ intermediate states with initial state of one massless
spin $1/2$ quark (or anti-quark) and one massless spin $1$ gluon.
The third distribution, eq.(\ref{J=1in2}), indicates a spin $1$ state in
spin $2$ intermediate state, where massless quark and anti-quark
couple with a spin $1$ state.

\begin{figure}[t]
\centering
\includegraphics[width=45mm]{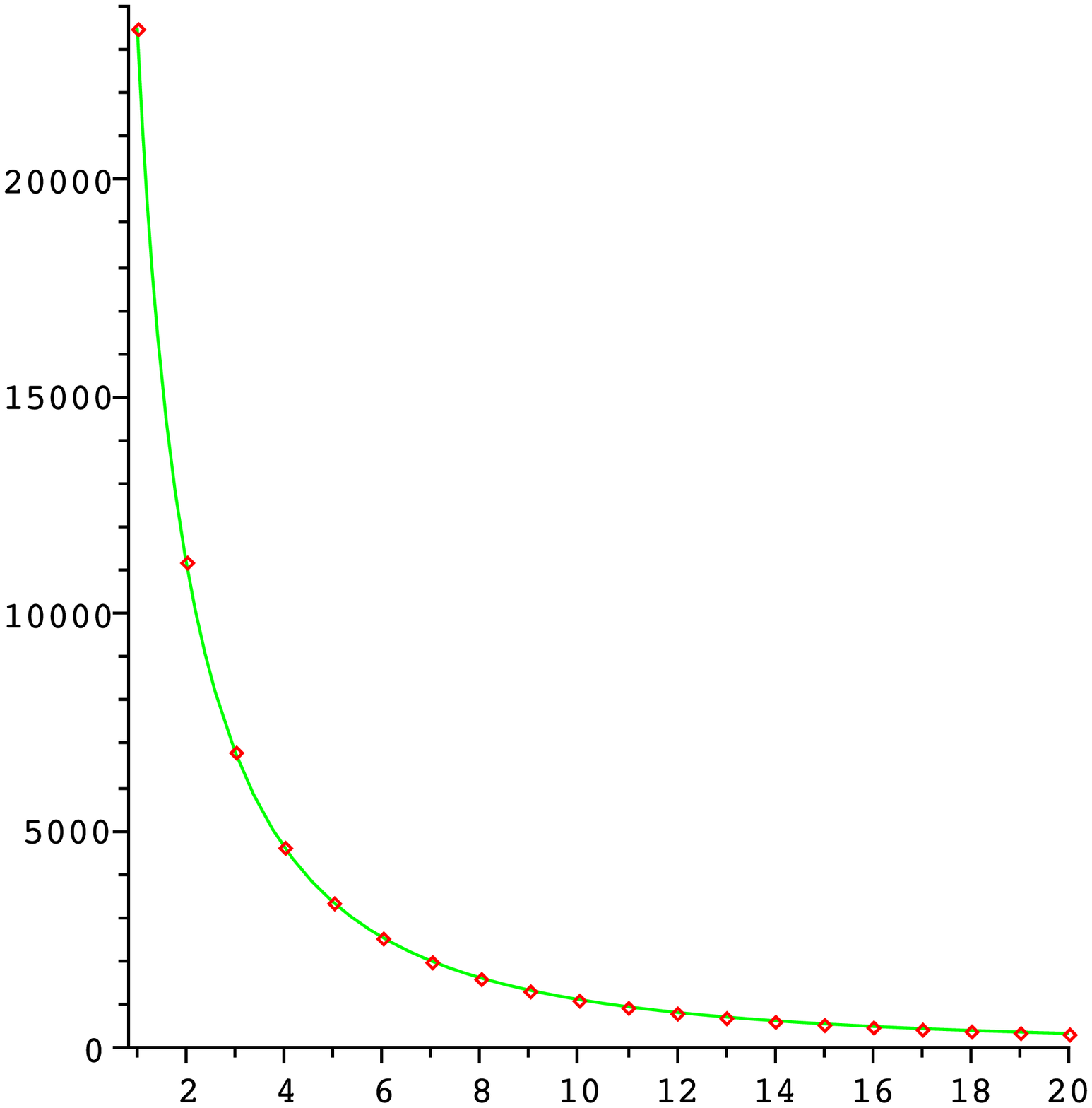}
\includegraphics[width=45mm]{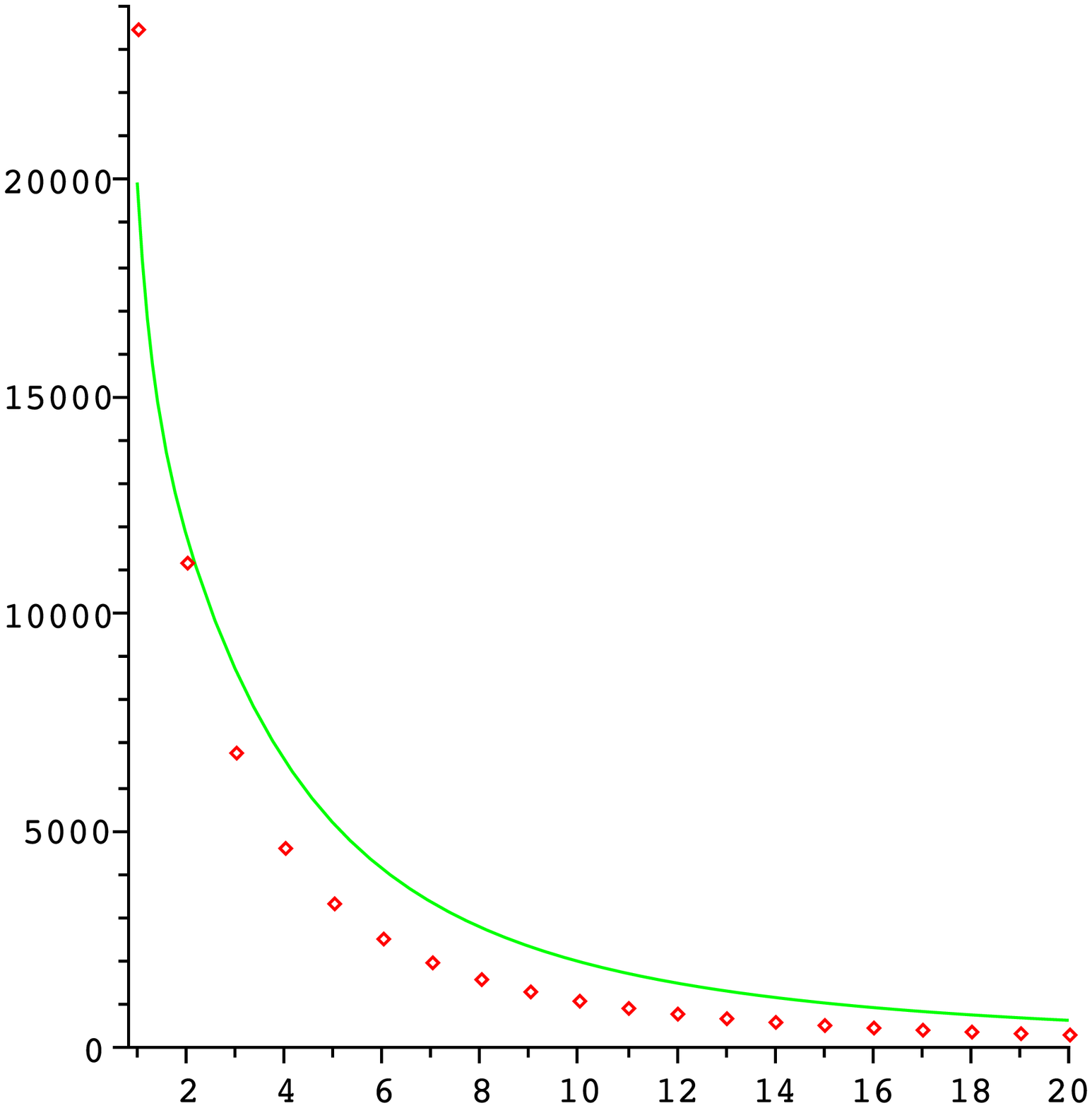}
\includegraphics[width=45mm]{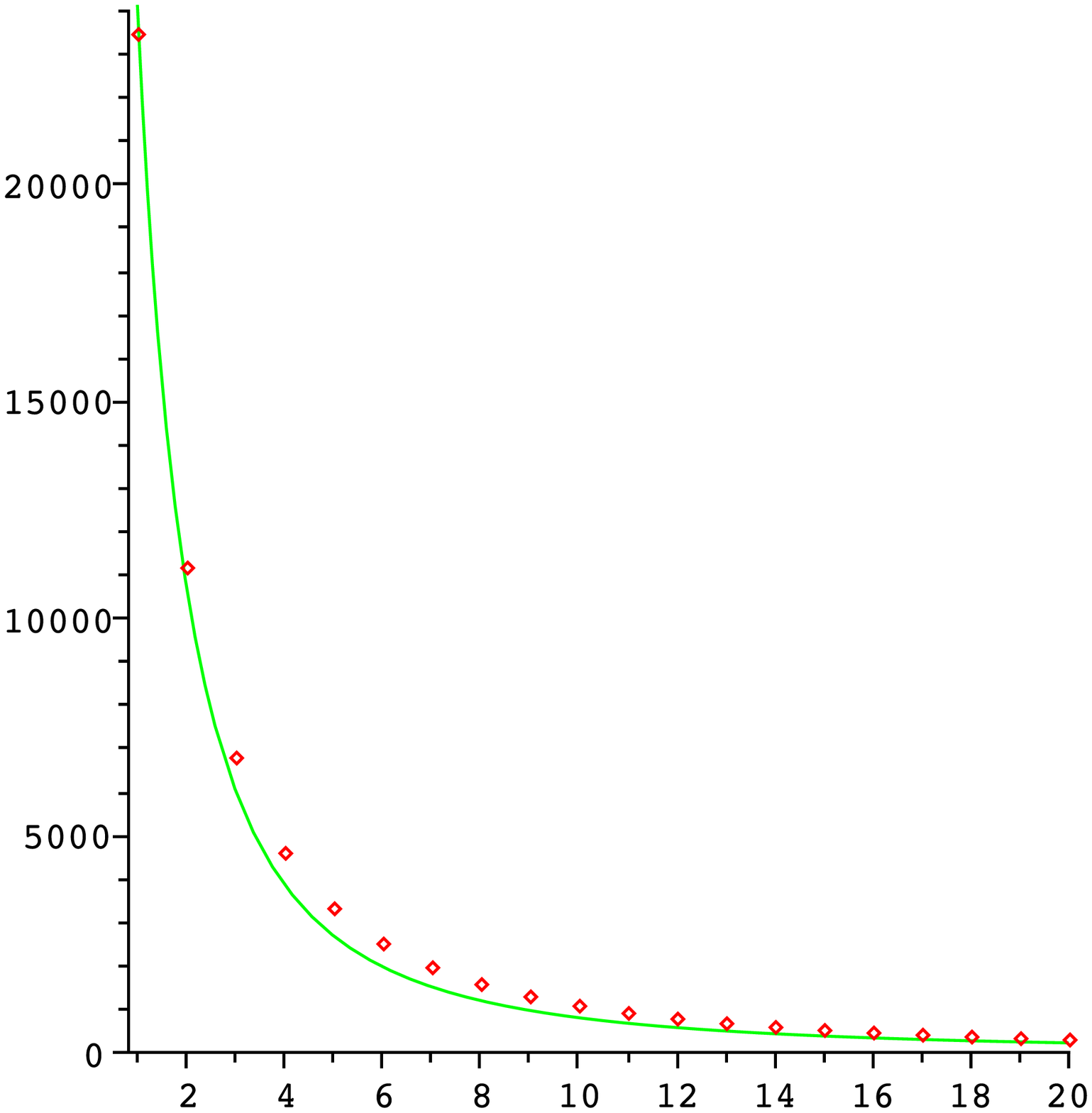}
\caption{
Three fits for ``on peak'' $\chi$-distribution ($M_s = 2$ TeV)
with eq.(\ref{J=1/2-3/2}). Red points indicate the distribution
by string excited stets and green lines indicate fit lines.
From left to right, fit with $A_2,B_2 \ne 0$ (both spin $1/2$ and
$3/2$ intermediate states), fit with $A_2=0$ (only spin $3/2$ intermediate
state) and fit with $B_2=0$ (only spin $1/2$ intermediate state).
}
\label{chi-fit}
\end{figure}

Fig.\ref{chi-fit} show various fits of ``on peak'' $\chi$-distribution
in case of $M_s=2$ TeV with eq.(\ref{J=1/2-3/2}).
Since the process $qg \rightarrow qg$ dominates ``on peak'' distributions,
it should be fit well with both non-vanishing $A_2$ and $B_2$. This is true
as it is shown in the first figure in Fig.\ref{chi-fit}. If this fit is better
than that the other two fits, assuming only spin $3/2$ intermediate state
or only spin $1/2$ intermediate state, we see that at least two degenerate
states contribute, which is the prediction of string models.
The second and the third figures in Fig.\ref{chi-fit} show the fits assuming
single intermediate state. We see that two-state fit is better than single
state fit. Since the actual situation depends on experimental precision and
uncertainties, the feasibility of this procedure should be considered with
detector simulations. Subtraction of QCD background is also a crucial issue.

It would be very difficult to distinguish small difference between
eq.(\ref{J=0-2}) and eq.(\ref{J=1/2-3/2}) experimentally.
Even though the process $gg \rightarrow gg$ dominates ``off peak''
distribution with $M_s=2$ TeV, the goodness of fit with eq.(\ref{J=0-2})
and that with eq.(\ref{J=1/2-3/2}) are almost the same.
If we could separate gluon jets and quark jets with a certain efficiency,
investigating both ``on peak'' and `off peak'' distributions could be
meaningful.

\section{Effects of the mass shift in gluon dijet events}
\label{gluon-dijet}

If it is possible to distinguish gluon jets and quark jets in a certain
efficiency, it may be possible to investigate the process $gg \rightarrow gg$
independently. This is an interesting process with contributions of four
string excited states as we can see in eq.(\ref{amp2-gg}). Though two
color octet states obtain small one-loop correction to their masses, the
masses of two color singlet states may obtain large one-loop corrections.
It is shown in ref.\cite{Antoniadis:2002cs} by explicit calculations in
string world-sheet theory that anomalous U$(1)$ gauge boson, which is
massless at tree level, obtain mass of the order of $0.1 \times M_s^2$ or larger
through the open string one-loop effect, or through the tree-level mixing
with closed string states. The same may happen for the states $C^*_{J=0}$
and $C^*_{J=2}$. In the next section, we will estimate the shift of mass
squared of $C^*_{J=2}$ is $\Delta m^2 = (3/\pi^3) M_s^2$ by explicit 
calculations.
Namely, the shift of mass squared of $C^*_{J=2}$ (and $C^*_{J=0}$) is
of the order of $10\%$ of $M_s^2$. This gives a non-trivial shape to
resonance structure by the process $gg \rightarrow gg$.

\begin{figure}[t]
\centering
\includegraphics[width=45mm]{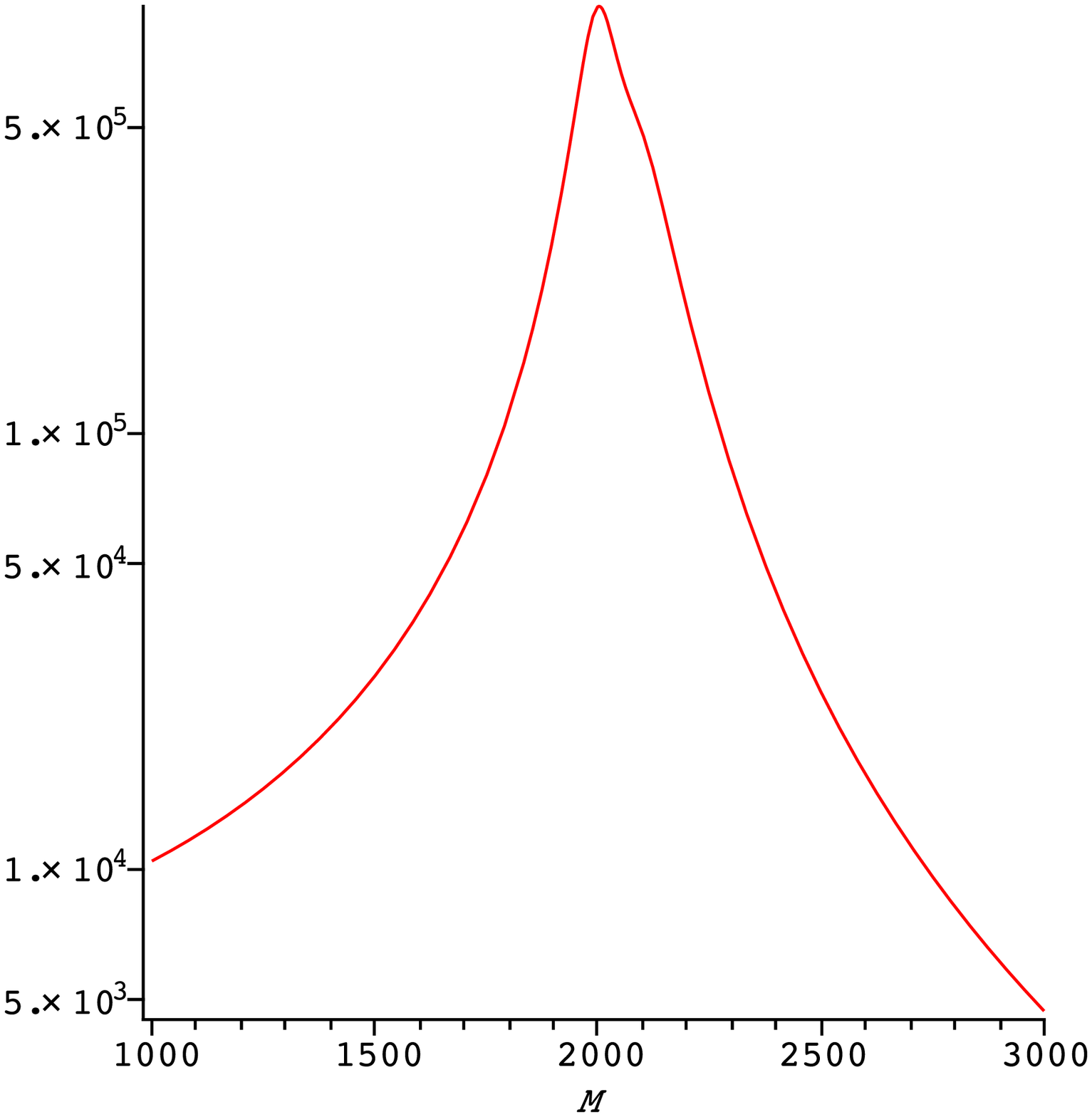}
\includegraphics[width=45mm]{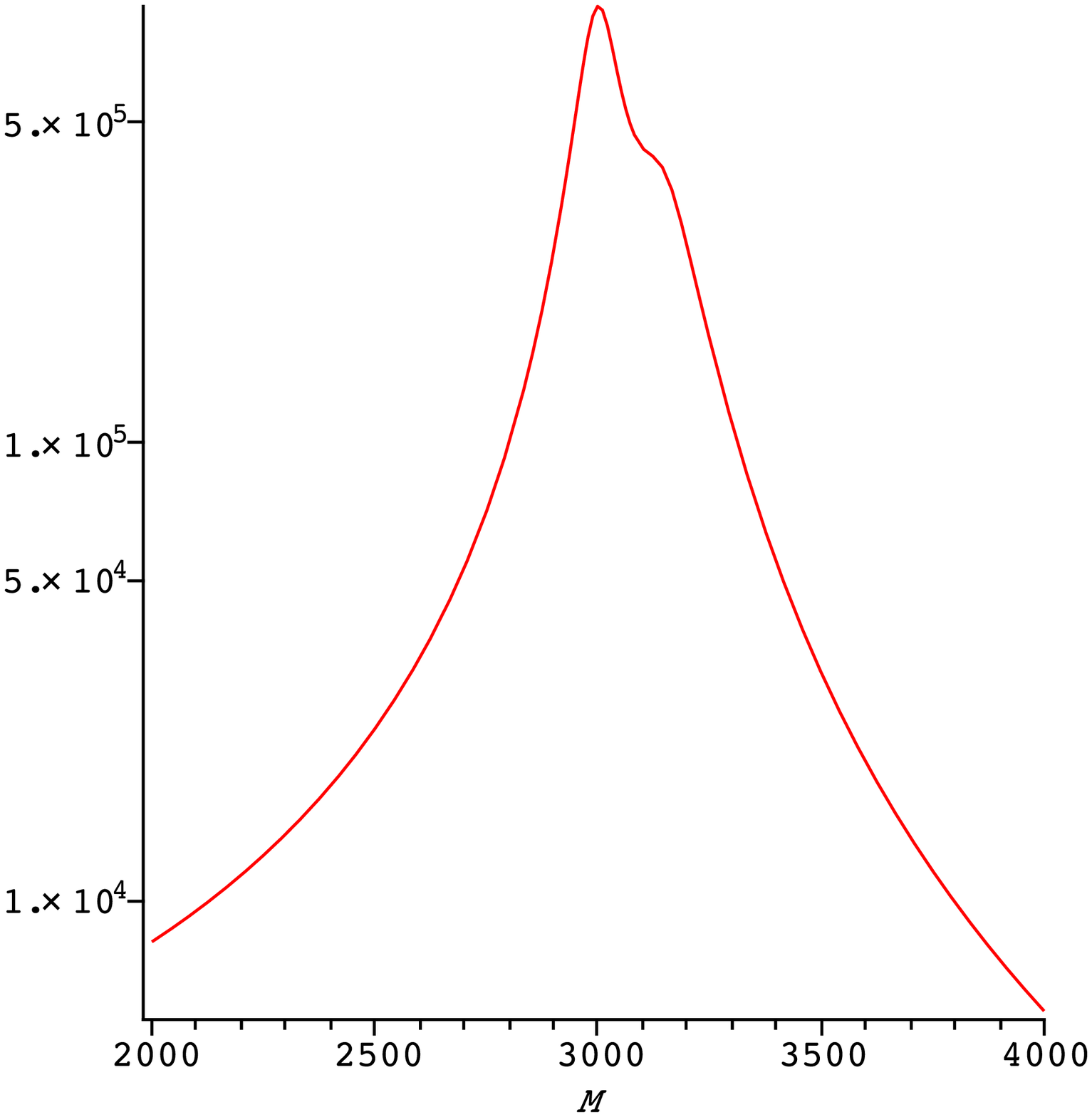}
\includegraphics[width=45mm]{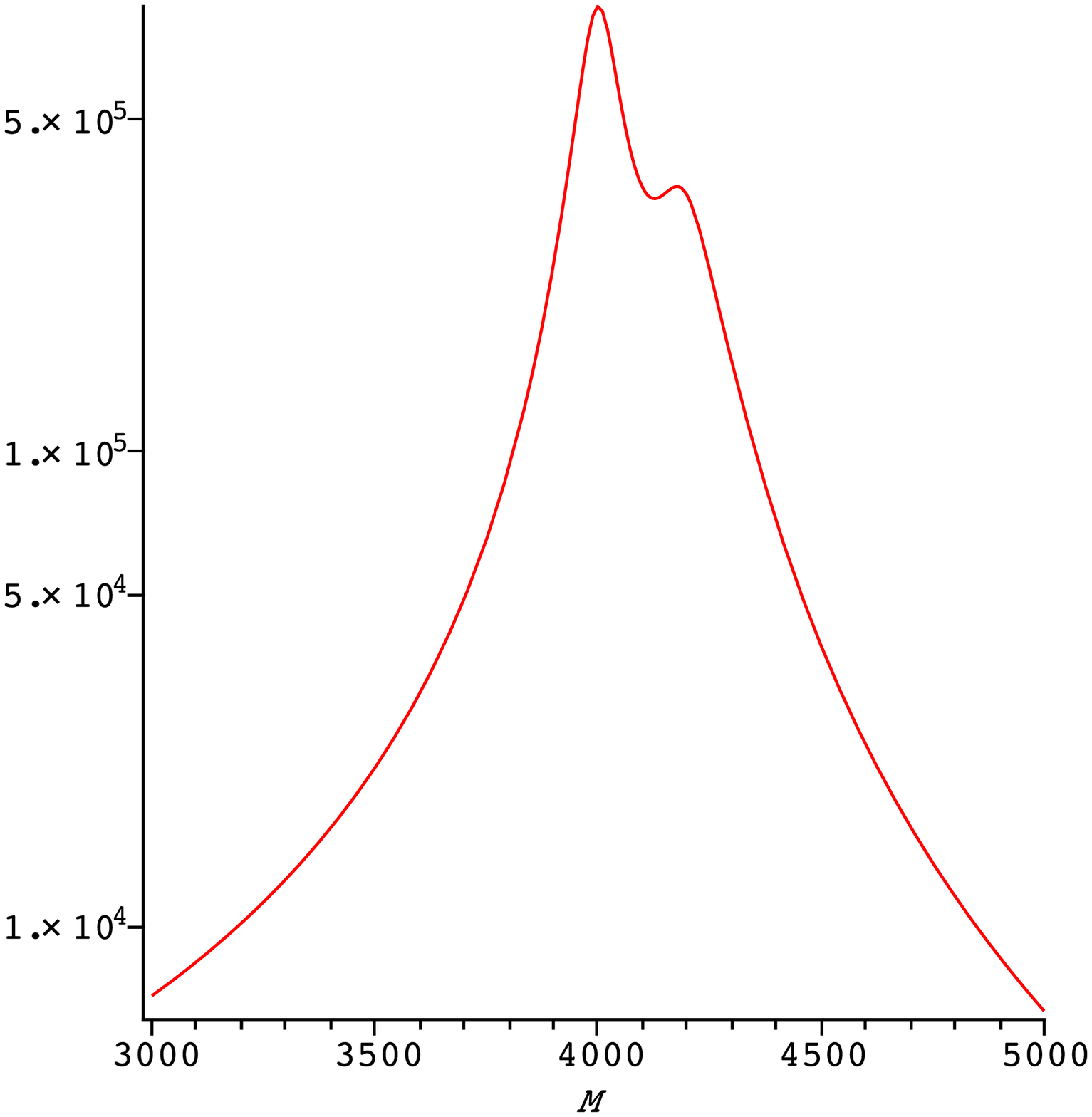}
\caption{
Total cross sections (unit [fb]) of the process $gg \rightarrow gg$
in ideal (unrealistic) gluon-gluon collider scanning $M=\sqrt{\hat{s}}$.
From left to right, $M_s=2,3$ and $4$ TeV.
}
\label{gg-sigma}
\end{figure}

\begin{figure}[t]
\centering
\includegraphics[width=45mm]{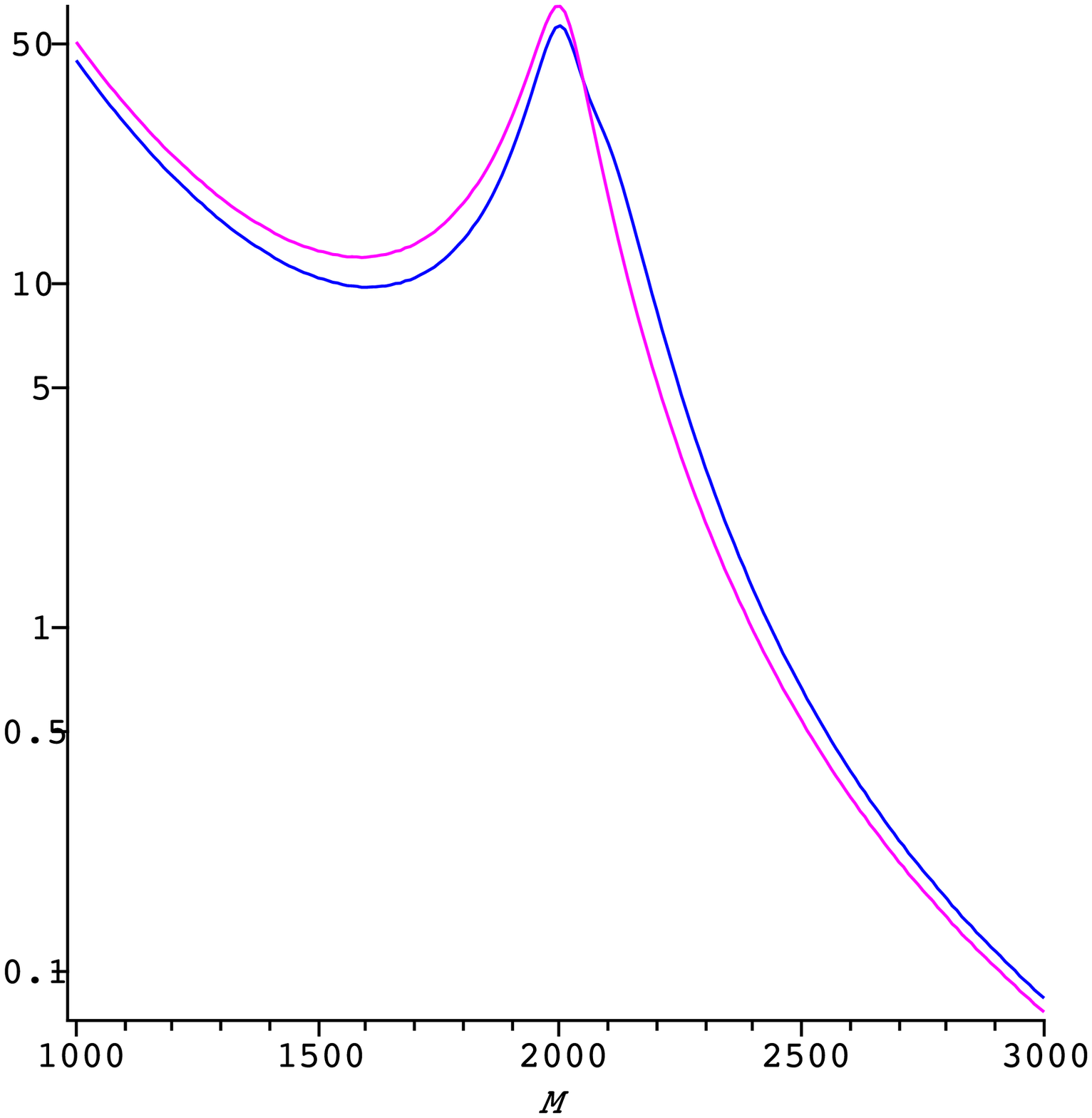}
\includegraphics[width=45mm]{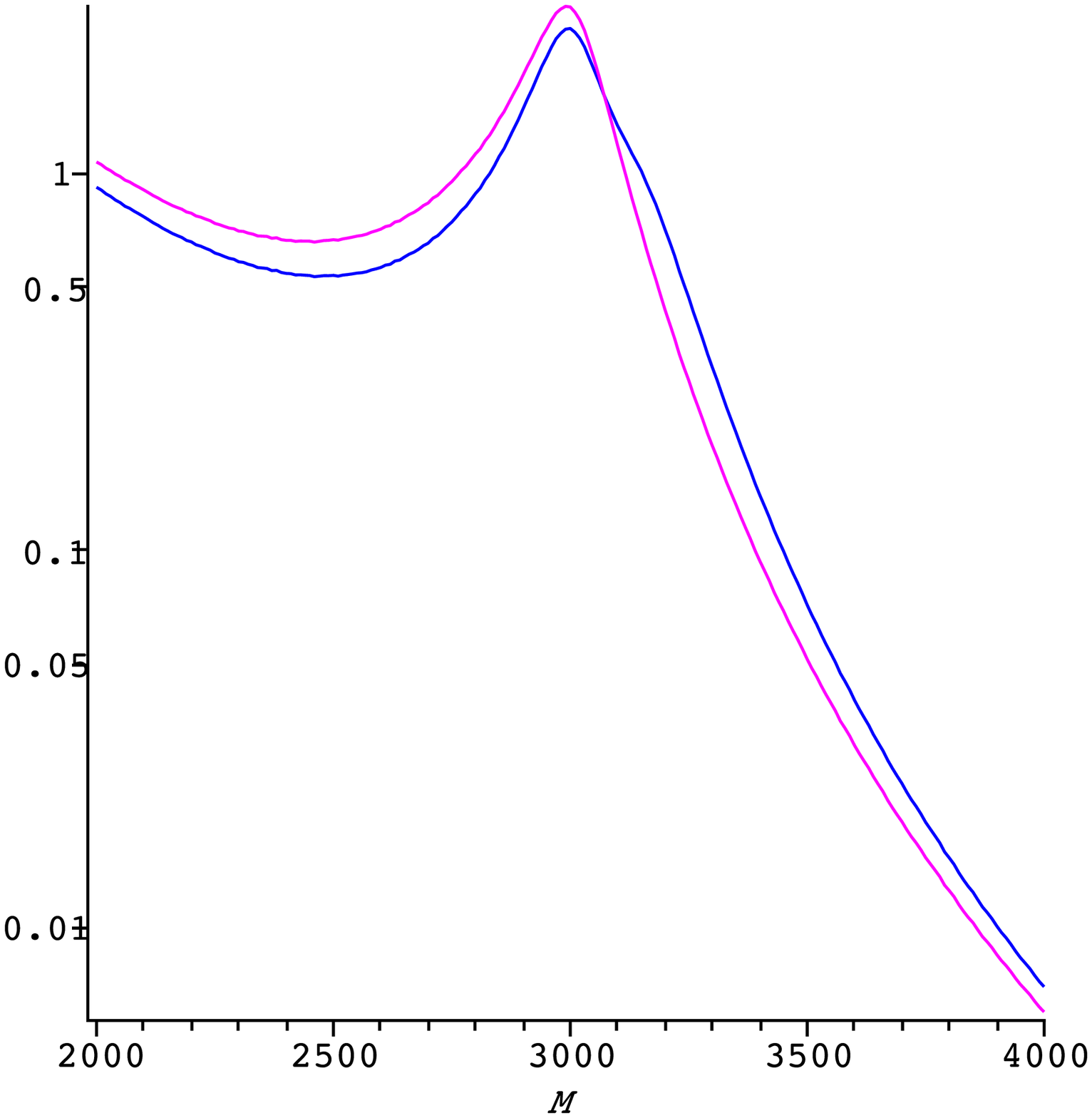}
\includegraphics[width=45mm]{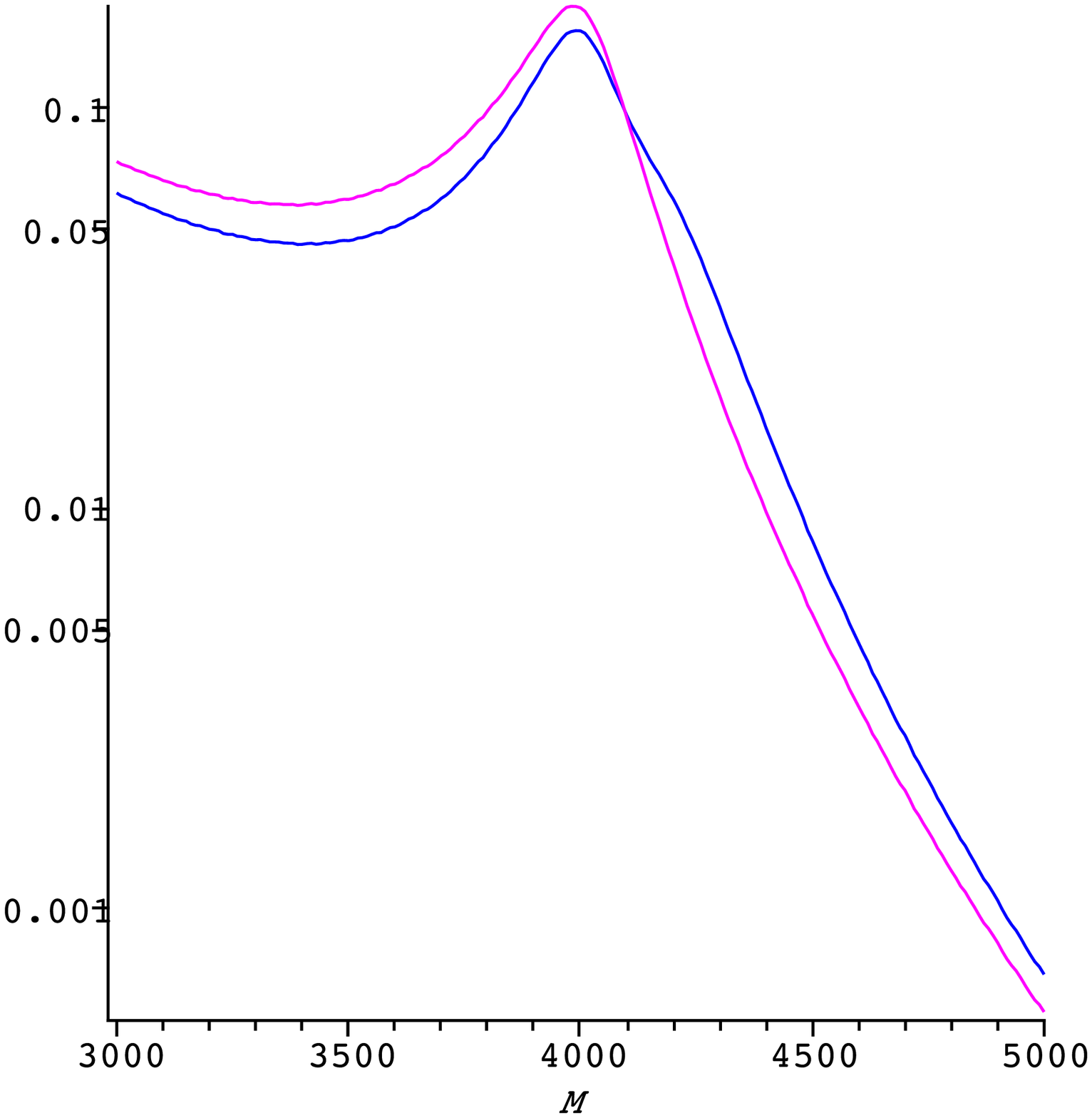}
\caption{
Dijet invariant mass distributions (unit [fb/GeV]) of the process
$gg \rightarrow gg$ without QCD background. The line with color blue
is that with mass shifts and the line with color magenta is that
without mass shifts. From left to right, $M_s=2,3$ and $4$ TeV.
}
\label{gg-dist}
\end{figure}

Fig.\ref{gg-sigma} show parton-level total cross sections of the process
$gg \rightarrow gg$ with mass shifts. We see that the shape of simple
Breit-Wigner type resonance is largely distorted especially for larger value
of $M_s$. In real observable, dijet invariant mass distributions, this
distortion is smeared as shown in Fig.\ref{gg-dist} (with rapidity cut
$\vert y_1 \vert$, $\vert y_2 \vert < 1$ and $\sqrt{s}=14$ TeV). We see that
the value of the left of the peak decreases and the value of the right of the
peak increases by the effect of mass shifts.

\begin{figure}[t]
\centering
\includegraphics[width=45mm]{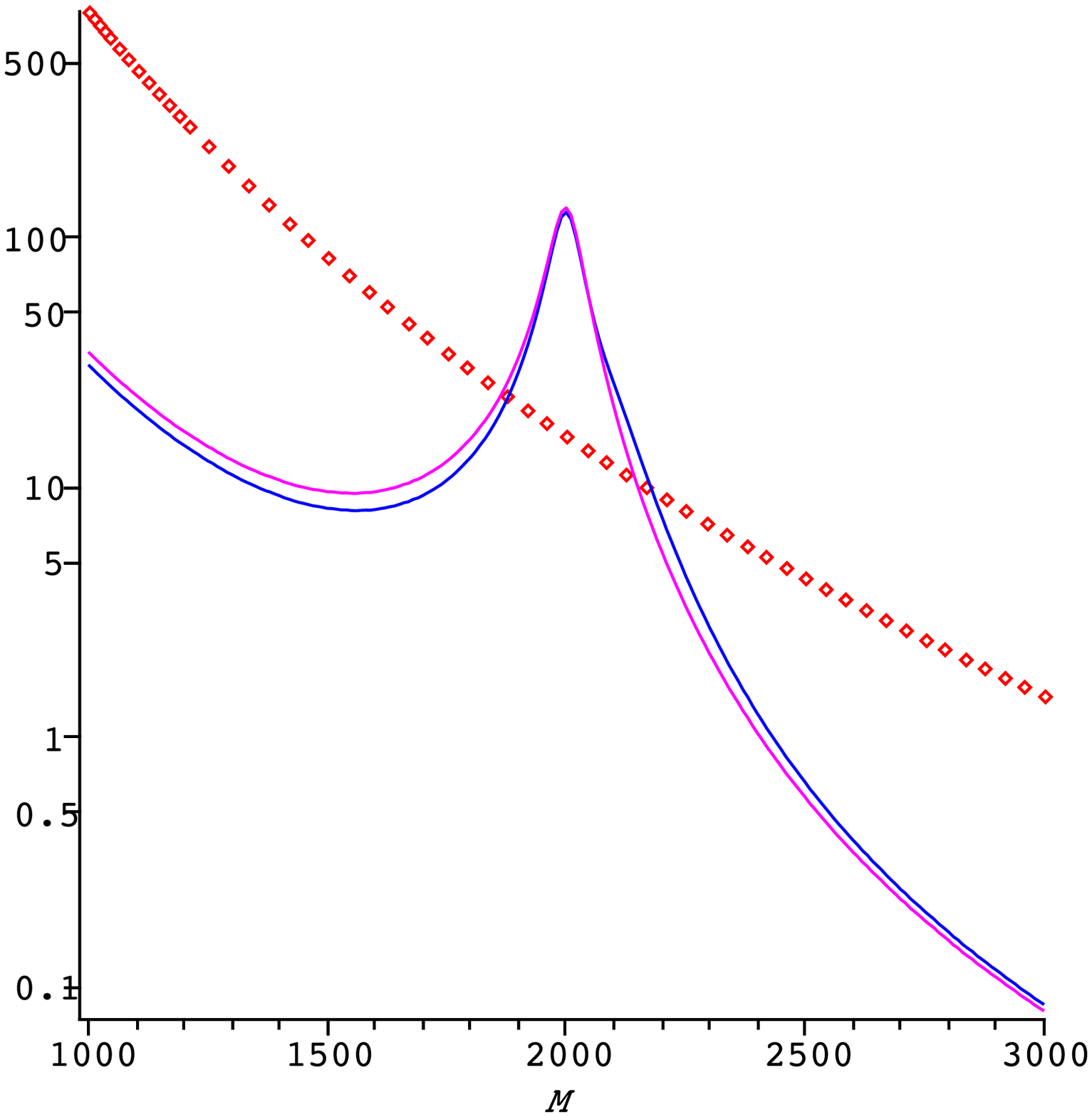}
\includegraphics[width=45mm]{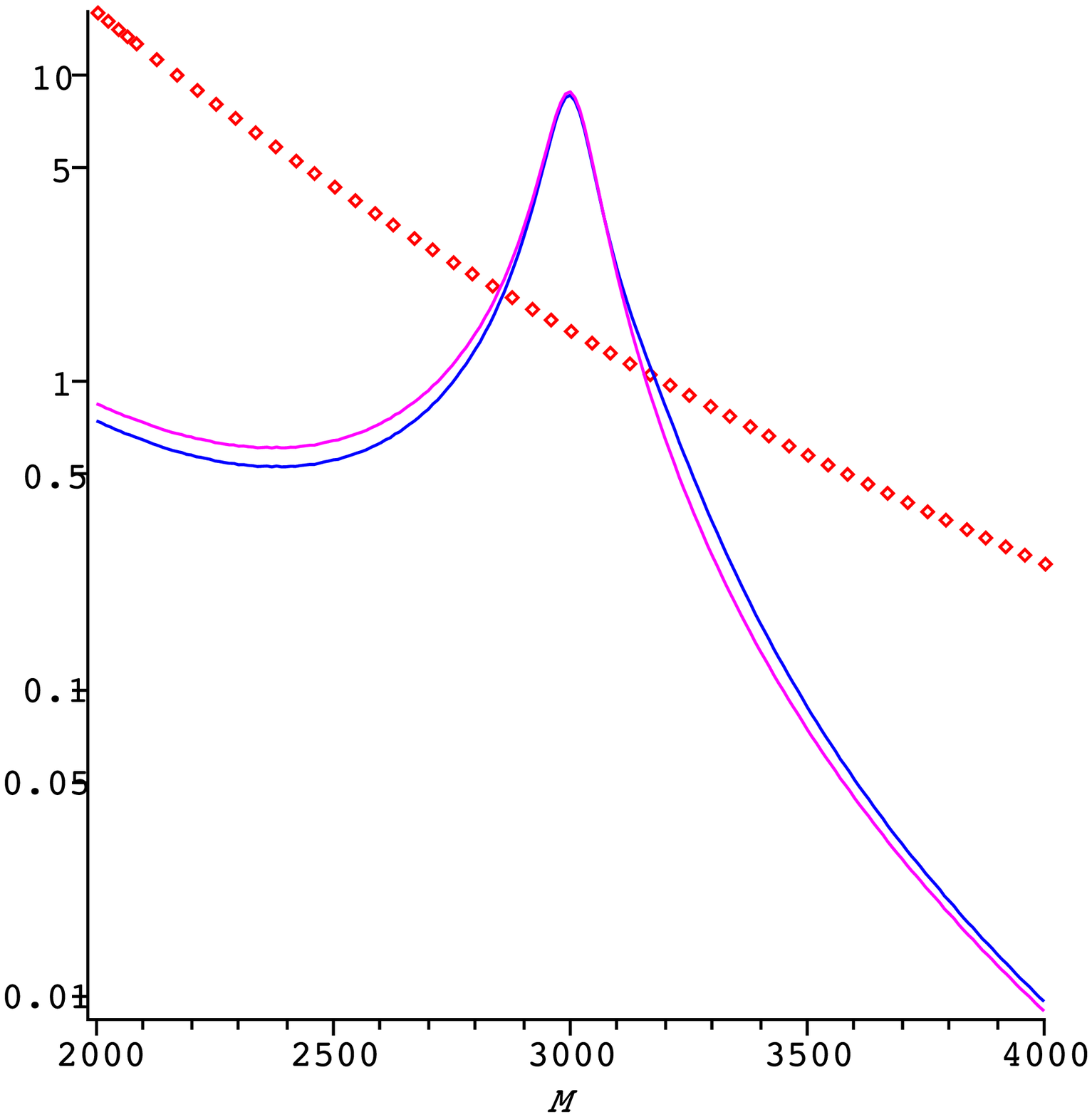}
\includegraphics[width=45mm]{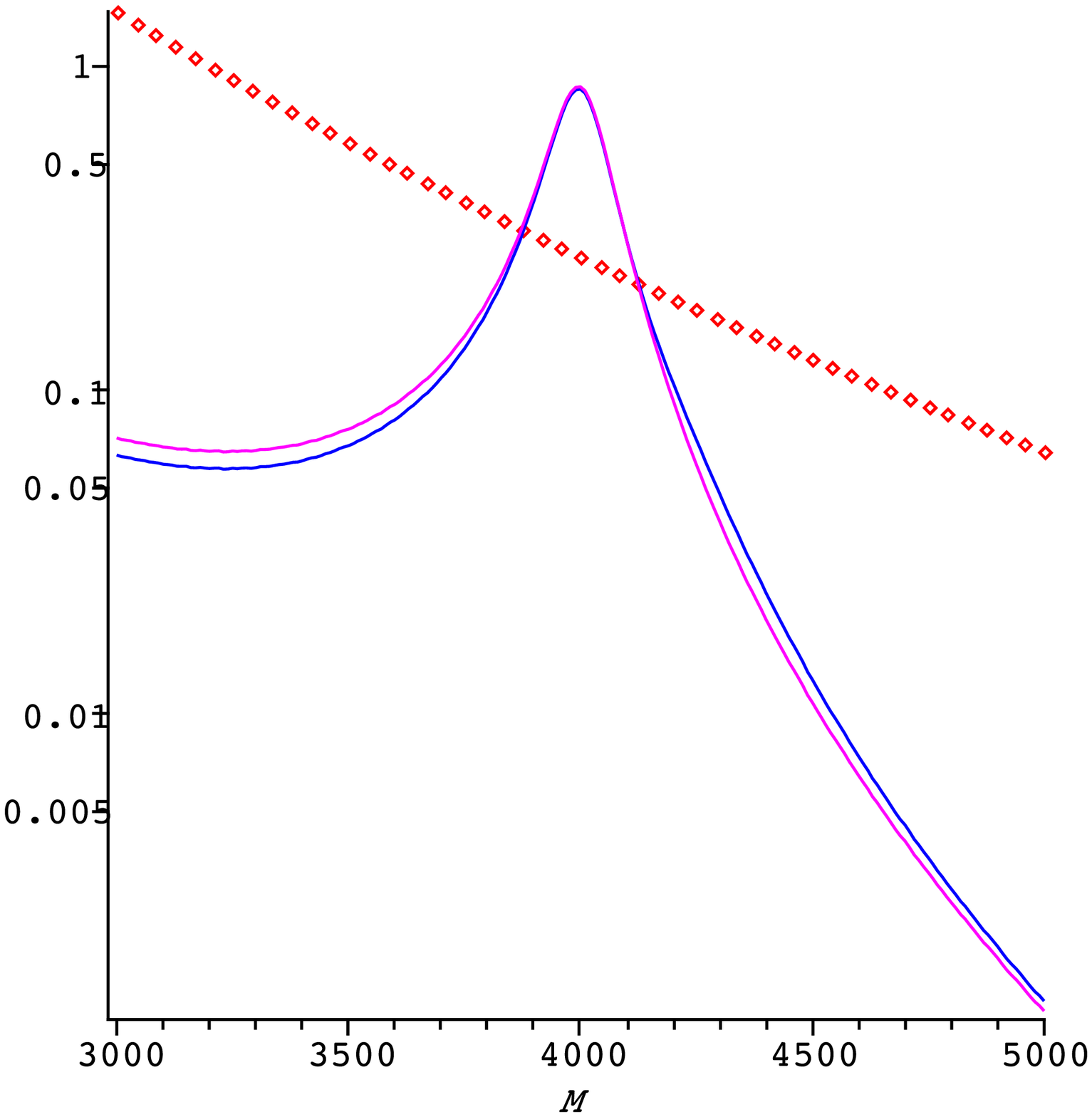}
\caption{
Dijet invariant mass distribution of the process $gg \rightarrow gg$ plus
$qg \rightarrow qg$ (unit [fb/GeV]). The line with color blue is that with mass
shifts and the line with color magenta is that without mass shifts. The points
with red denote QCD backgrounds. From left to right, $M_s=2,3,4$ TeV.
}
\label{dist-obs}
\end{figure}

In actual experiments, we can not escape from the contamination due to dominant
process $qg \rightarrow qg$, even if a certain level of gluon jets selection
might be possible. Fig.\ref{dist-obs} show dijet invariant mass distributions
including the process $qg \rightarrow qg$ with reduction of $1/3$ and the
process $gg \rightarrow gg$ with reduction of $2/3$. Assuming the subtraction
of QCD background, we can consider the following ratio of cumulative cross
sections as a measure of the existence of mass shifts or distortion from
simple Breit-Wigner type shape.
\begin{equation}
 R \equiv {{\sigma_{\rm left}-\sigma_{\rm right}}
           \over
           {\sigma_{\rm left}+\sigma_{\rm right}}},
\end{equation}
 where
\begin{equation}
 \sigma_{\rm left}
  \equiv \int_{M_s-\Delta M}^{M_s} dM {{d\sigma} \over {dM}},
\qquad
 \sigma_{\rm right}
  \equiv \int_{M_s}^{M_s+\Delta M} dM {{d\sigma} \over {dM}}.
\end{equation}
The values of this ratio in case of $\Delta M = 1000$ GeV are
\begin{eqnarray}
 &
 R_{\Delta m^2 \ne 0} \simeq 0.26,
 \quad
 R_{\Delta m^2 = 0} \simeq 0.31
 \quad
 (R_{\Delta m^2 \ne 0}/R_{\Delta ^2 = 0} \simeq 0.83)
 \quad \mbox{for $M_s = 2$ TeV},
 &
\\
 &
 R_{\Delta m^2 \ne 0} \simeq 0.18,
 \quad
 R_{\Delta m^2 = 0} \simeq 0.22
 \quad
 (R_{\Delta m^2 \ne 0}/R_{\Delta ^2 = 0} \simeq 0.80)
 \quad \mbox{for $M_s = 3$ TeV},
 &
\\
 &
 R_{\Delta m^2 \ne 0} \simeq 0.19,
 \quad
 R_{\Delta m^2 = 0} \simeq 0.22
 \quad
 (R_{\Delta m^2 \ne 0}/R_{\Delta ^2 = 0} \simeq 0.85)
 \quad \mbox{for $M_s = 4$ TeV}.
 &
\end{eqnarray}
The number of events ${\cal L}\sigma_{\rm left}$
with ${\cal L}=100 \ {\rm fb}^{-1}$ are of the order of
$10^5$, $10^4$ and $10^3$ for $M_s=2,3$ and $4$ TeV, respectively.
In case of small $M_s \sim 2$ TeV, the selection of gluon jets is not
so crucial, since the process $gg \rightarrow gg$ dominates in left region.
In case of larger $M_s$ the selection becomes more important.
The precise determination of QCD background is very important not to generate
systematic unbalance in left and right regions by the subtraction.


\section{Calculation of the mass shift in String Theory}
\label{mass-shift}

The one-loop corrections to the masses of first string excited states
can be calculated in string world-sheet theory. We calculate one-loop
mass correction to the state $C^*_{J=2}$ by investigating its two point
amplitude along the line of ref.\cite{Antoniadis:2002cs} in which
one-loop mass of anomalous U$(1)$ gauge boson is explicitly calculated.
It is assumed that color gauge symmetry is on a stack of D3-brane.
The two point amplitude of $C^*_{J=2}$ can be written as follows.
\begin{equation}
 A_2 = \int_0^\infty {{dt} \over {2t}} \int_0^t dx \
       \langle V_1(z_1) V_2(z_2) \rangle_t,
\end{equation}
 where $z_1 = \exp(-x)$ and $z_2 = \exp(i\pi)$, and $\langle \cdots \rangle_t$
 denotes correlation functions on cylinder with modulus $t$.
The operators $V_1(z_1)$ and $V_2(z_2)$ are vertex operators with picture $0$
 of $C^*_{J=2}$ with momenta $k_1$ and $k_2$ (both incoming) and polarizations
 $e^{(1)}_{\mu\nu}$ and $e^{(2)}_{\mu\nu}$ (traceless antisymmetric, and
 $k^\mu e_{\mu\nu}=0$) in four dimensional spacetime.
\begin{eqnarray}
 V_1(z_1) &=& {1 \over \sqrt{2 \alpha'}} \ T_{C^*_{J=2}} \
            e^{(1)}_{\mu\nu}
            \left\{
             i \partial X^\mu i \partial X^\nu
             + 2 \alpha' \left(
                          (k_1 \cdot \psi) i \partial X^\mu \psi^\nu
                          + \partial \psi^\mu \psi^\nu
                         \right)
            \right\} e^{i k_1 \cdot X},
\\
 V_2(z_2) &=& {1 \over \sqrt{2 \alpha'}} \ T_{C^*_{J=2}} \
            e^{(2)}_{\mu\nu}
            \left\{
             i \partial X^\mu i \partial X^\nu
             + 2 \alpha' \left(
                          (k_2 \cdot \psi) i \partial X^\mu \psi^\nu
                          + \partial \psi^\mu \psi^\nu
                         \right)
            \right\} e^{i k_2 \cdot X},
\end{eqnarray}
 where $T_{C^*_{J=2}} = {\rm diag}(1/\sqrt{6},1/\sqrt{6},1/\sqrt{6})$
 is Chan-Paton matrix for the state $C^*_{J=2}$.
We consider only the non-planar diagram, which include the effect of
mixing between closed string states and open string states.

The amplitude can be explicitly written as follows.
\begin{eqnarray}
 A_2 &=& ({\rm tr}(T_{C^*_{J=2}}))^2 \
         {1 \over {2 \alpha'}} \int_0^\infty {{dt} \over {2t}} \int_0^t dx \
         2 \ e^{(1)}_{\mu\nu} e^{(2)}{}^{\mu\nu} \
         {1 \over 2} \sum_{\alpha,\beta = 0,1} (-1)^\alpha
         \langle e^{ik_1 \cdot X} e^{ik_2 \cdot X} \rangle
\nonumber\\
 && \qquad \times
    \Bigg\{
     (2 \alpha')^2 \langle \partial \psi \psi \rangle
                     \langle \psi \partial \psi \rangle
     + \langle \psi \psi \rangle^2
       \langle \partial X X \rangle \langle X \partial X \rangle
    \Bigg\}
    Z^{\alpha\beta}(t),
\end{eqnarray}
 where 
\begin{eqnarray}
 (2 \alpha')^2 \langle \partial \psi \psi \rangle
               \langle \psi \partial \psi \rangle
 &=& {{(2 \alpha')^2} \over {8 z_1}} \cdot
     {{({\cal P}'(z_1+1))^2}
      \over
      {{\cal P}(z_1+1)
       + 4 \pi i \partial_\tau
         \ln \left(
          \theta\left[\begin{array}{c} \alpha/2 \\ \beta/2 \end{array}\right]
            (0\vert\tau) \bigg/ \eta(\tau) \right)}}
\\
 \langle \psi \psi \rangle^2
 \langle \partial X X \rangle \langle X \partial X \rangle
 &=& - 2 \pi i \partial_\tau
     \ln \left(
          \theta\left[\begin{array}{c} \alpha/2 \\ \beta/2 \end{array}\right]
           (0\vert\tau)
            \Big/ \eta(\tau)
           \right)
\\
 &&
           \times
           {1 \over {z_1}}
           \left({{\alpha'} \over {2\pi}}\right)^2
           \left\{{{\theta_1'((ix-\pi)/2\pi \vert \tau)}
                   \over
                   {\theta_1((ix-\pi)/2\pi \vert \tau)}}
                   - {{ix} \over {\tau_2}}
           \right\}
           \left\{{{\theta_1'((\pi-ix)/2\pi \vert \tau)}
                   \over
                   {\theta_1((\pi-ix)/2\pi \vert \tau)}}
                   + {{ix} \over {\tau_2}}
           \right\},
\nonumber\\
\langle e^{ik_1 \cdot X} e^{ik_2 \cdot X} \rangle
 &=& \left(
    {{2 \pi (\eta(\tau))^3}
     \over
     {\theta_2 (ix/2 \pi \vert \tau)}}
    e^{{\pi \over {(2\pi)^2}}{{x^2} \over {\tau_2}}}
   \right)^{2 \alpha' k_1 \cdot k_2},
\label{propagation}
\end{eqnarray}
 and
\begin{equation}
 Z^{\alpha\beta}(t)
 = (-1)^{(1-\alpha)\beta}
   {{i V_4} \over {(\sqrt{8 \pi^2 \alpha' t})^4}}
   {1 \over {(\eta(\tau))^8}}
   \left(
    \theta\left[\begin{array}{c} \alpha/2 \\ \beta/2 \end{array}\right]
     (0\vert\tau)
     \Big/ \eta(\tau)
   \right)^4.
\end{equation}
Here, $\tau = \tau_1 + i \tau_2$ with $\tau_1=0$ and $\tau_2 = t$,
$V_4$ is the volume of four dimensional spacetime,
${\cal P}$ is the Weierstrass function and ${\cal P}'$ is its derivative,
$\eta(\tau)$ is the Dedekind function, and
\begin{equation}
 \theta\left[\begin{array}{c} a \\ b \end{array}\right] (z\vert\tau)
 \equiv e^{2 \pi i a (z+b)} q^{a^2/2} \prod_{n=1}^\infty (1-q^n)
        \prod_{m=1}^\infty (1 + q^{m+a-1/2} e^{2 \pi i (z+b)})
                           (1 + q^{m-a-1/2} e^{-2 \pi i (z+b)})
\end{equation}
 is the $\theta$-function ($q=e^{2 \pi i \tau}$) with $\theta_1(z \vert \tau)$
 is the case of $a=b=1/2$ and
 $\theta_1'(z \vert \tau) = \partial \theta_1(z \vert \tau) / \partial z$,
 and $\theta_2(z \vert \tau)$ is the case of $a=1/2$, $b=0$.

We can go from open string one-loop picture to closed string tree picture
by changing modulus variable from $t$ to $s=\pi/t$ and doing modular
transformation. As an order estimate, we consider the asymptotic form
in the limit of $s \rightarrow \infty$. Then the two point amplitude becomes
\begin{equation}
 A_2 \sim -i V_4 \ {1 \over 4} e^{(1)}_{\mu\nu} e^{(2)}{}^{\mu\nu} \
     ({\rm tr}(T_{C^*_{J=2}}))^2
     {1 \over {2\alpha'}} 
     {{16} \over {(4\pi)^3}}
     \int_0^\infty ds \ s^2 \ e^{-s/2}.
\end{equation}
The relations $k_2 = -k_1$ and $k_1 \cdot k_1 = - 1/\alpha' = -M_s^2$
(with different metric signature from that in previous sections)
have been used. The factor $e^{-s/2}$ comes from eq.(\ref{propagation})
indicating tree level propagation of massive closed string states.
Then the order of the mass shift becomes
\begin{equation}
 \Delta m^2
  \sim ({\rm tr}(T_{C^*_{J=2}}))^2
       {1 \over {2\alpha'}} 
       {{16} \over {(4\pi)^3}}
       \int_0^\infty ds \ s^2 \ e^{-s/2}
    = {3 \over {\pi^3}} M_s^2.
\end{equation}
This result means that the mass shift about $10\%$ of $M_s^2$ is expected
for color singlet open string states $C^*_{J=2}$ (and $C^*_{J=0}$).

\section{Conclusions}
\label{conclusions}

It has been stressed that a distinguished point of low-scale string models
from the other ``new physics'', like quark compositeness, on the resonance
in dijet invariant mass distribution is that a resonance consists many
resonances by many degenerate intermediate states with different spins.
Therefore, to discover or exclude low-scale string models, it is very
important to find the procedures to see the ``structure'' of resonance.

It has been proposed two procedures. One is the analysis of angular
distributions ($\chi$-distributions), and the other is looking for
the distortion of the resonance shape due to the mass shifts in string
excited states. Further detailed analysis including detector simulation
is necessary to clarify real feasibility of these procedures at the LHC.

\begin{figure}[t]
\centering
\includegraphics[width=45mm]{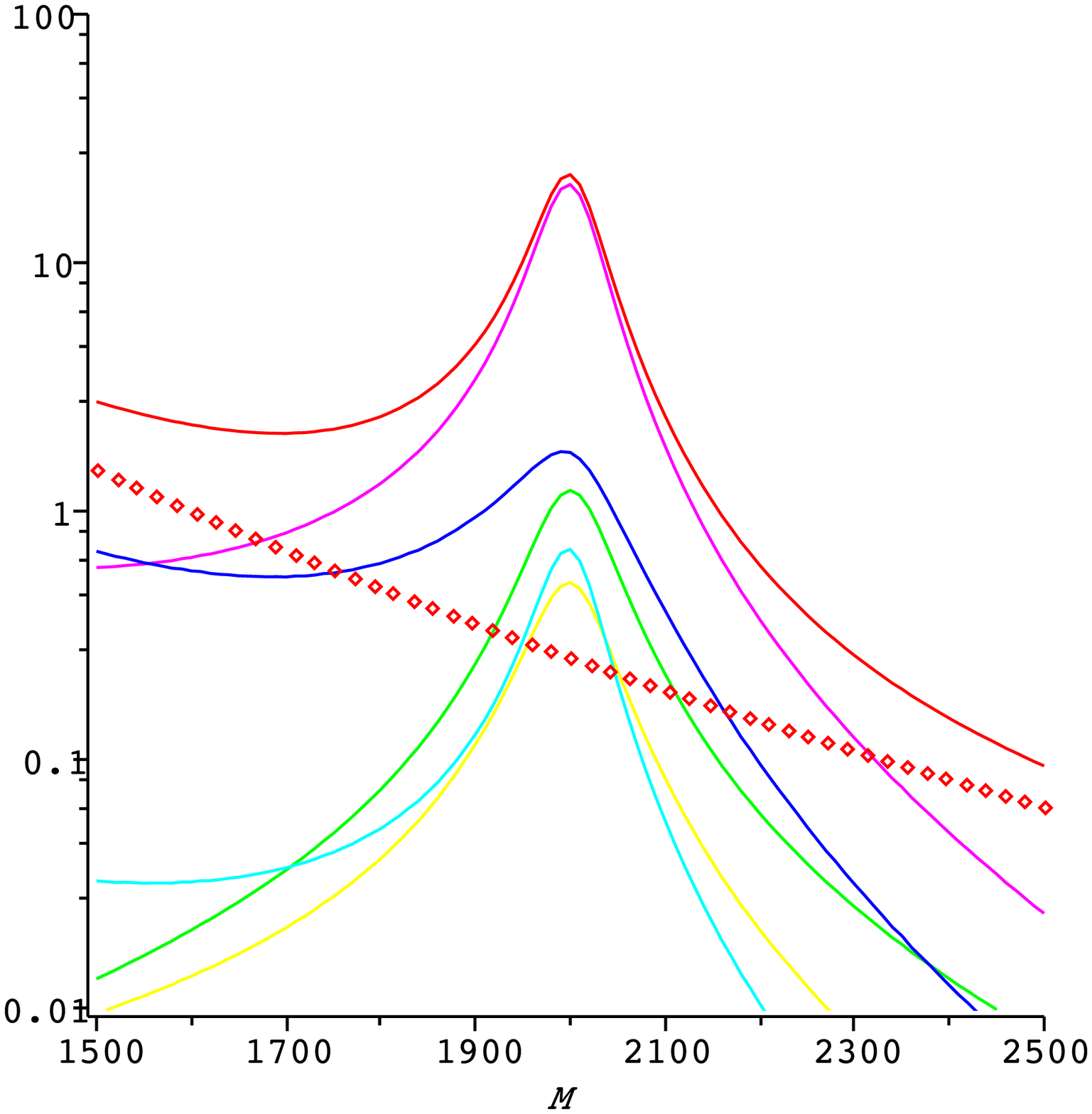}
\includegraphics[width=45mm]{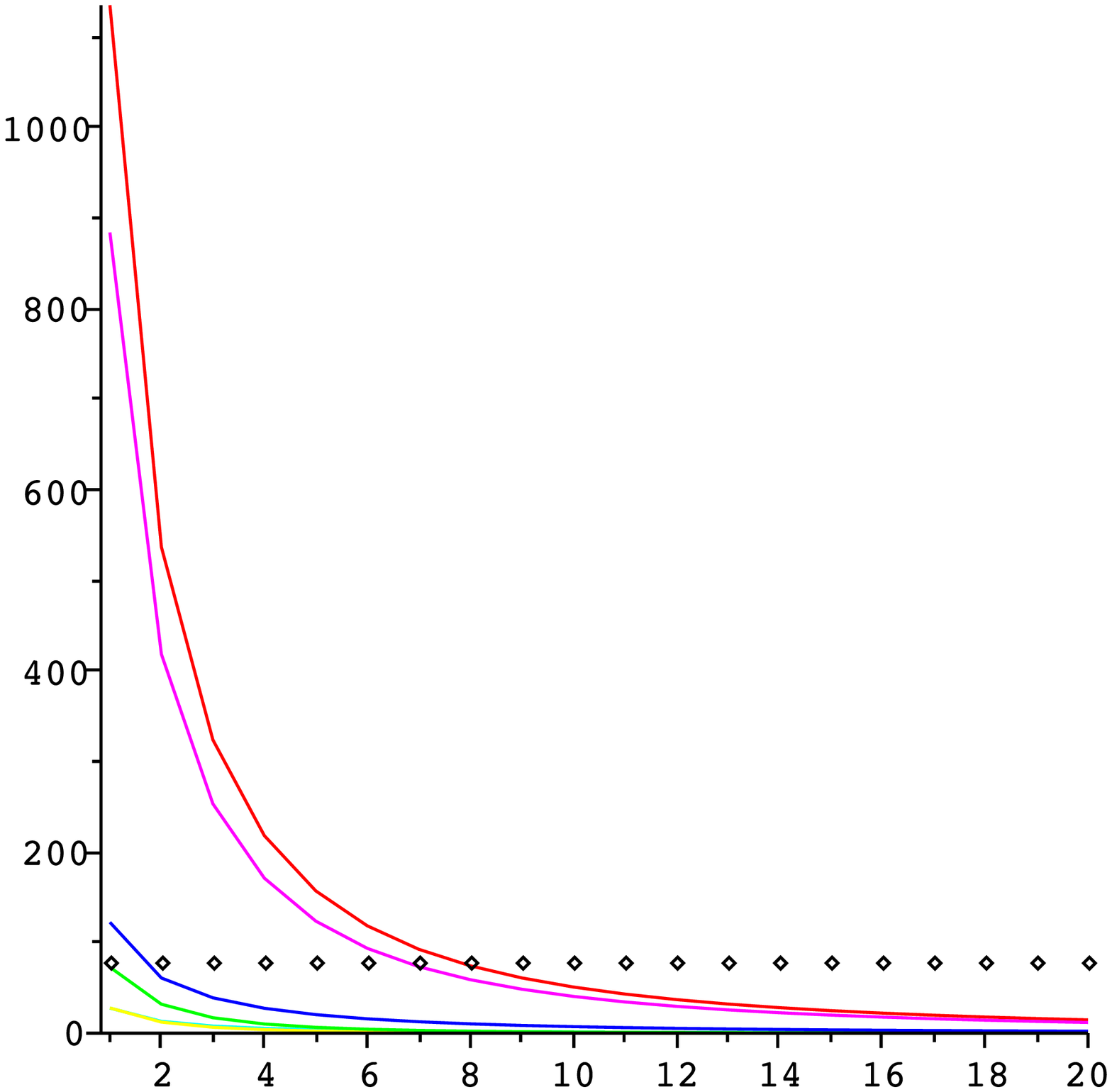}
\includegraphics[width=45mm]{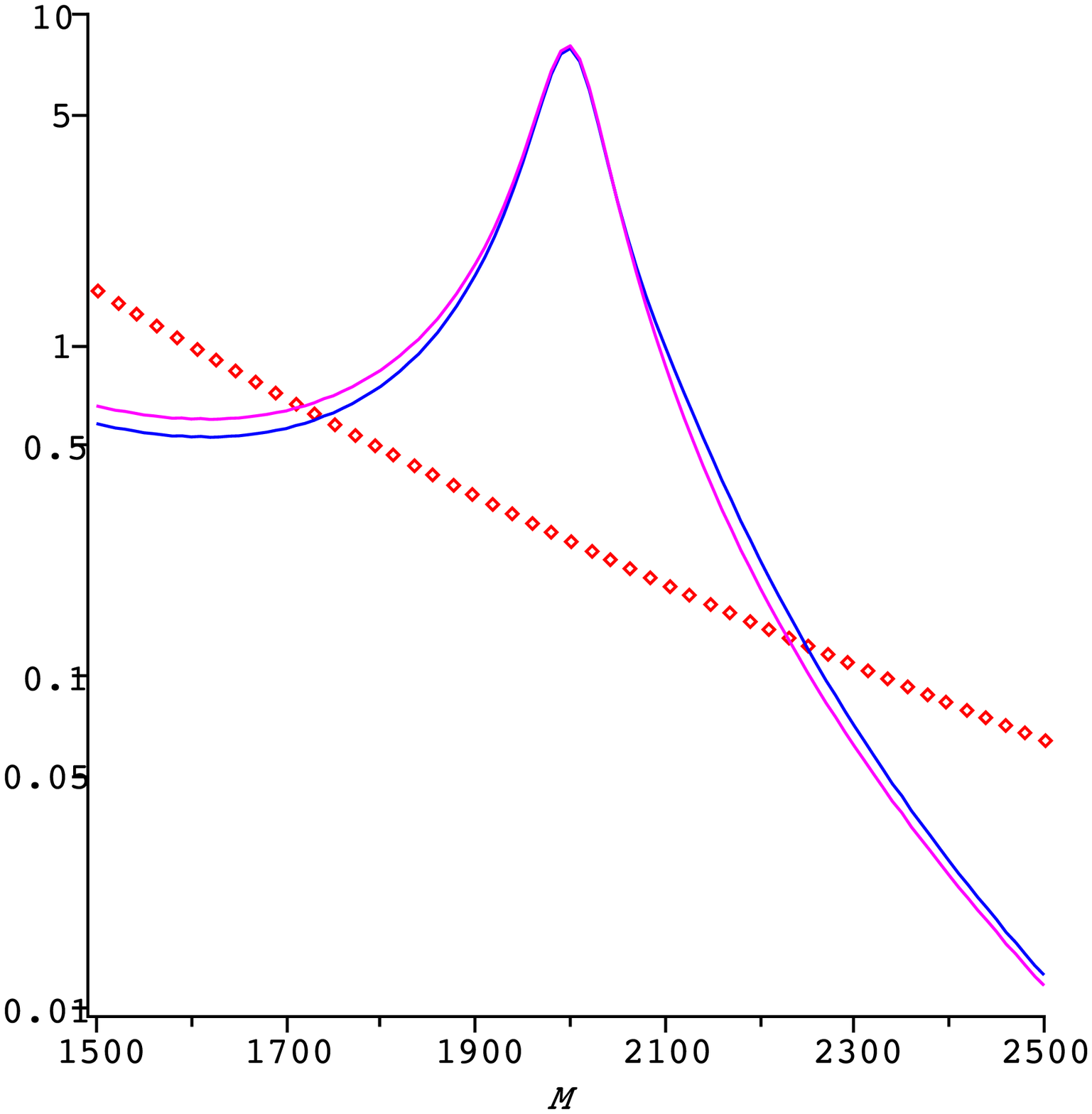}
\caption{
Plots in case of center of mass energy $7$ TeV. From left to right, dijet
invariant mass distribution (the same plot of Fig.\ref{fig:resonances-all}), 
``on peak''$\chi$-distribution (the same plot of Fig.\ref{chi-separate-on}),
and dijet invariant mass distribution with mass shift (the same plot of
Fig.\ref{dist-obs}).
}
\label{plots-7TeV}
\end{figure}

The center of mass energy of $pp$ collision $\sqrt{s}=14$ TeV has been
assumed in all the analysis in previous sections. It is planned that
the LHC will operate at $\sqrt{s}=7$ TeV until obtaining integrated
luminosity $1 \, {\rm fb}^{-1}$. In Fig.\ref{plots-7TeV} three plots
assuming $\sqrt{s}=7$ TeV for the case $M_s=2$ TeV are given.
In view of the number of events, the feasibility of the analysis seems
to be similar to the case of $M_s = 3$ TeV with $\sqrt{s}=14$ TeV.

\section*{Acknowledgments}

The author would like to thank Koji Terashi for useful information.

\end{document}